\documentclass[twocolumn,showkeys]{revtex4}

\usepackage{amssymb,amsfonts,amsmath}
\usepackage{graphicx}

\begin{document}













\title{Spatio-temporal correlations can drastically change the response of a MAPK pathway}

\author{Koichi Takahashi}
\email[E-mail:]{ktakahashi@riken.jp}
\affiliation{Advanced Sciences Institute, RIKEN,
    1-7-22 Suehirocho, Tsurumi, Yokohama, 230-0045, Japan}
\affiliation{The Molecular Sciences Institute, Berkeley, 2168 Shattuck
    Ave., Berkeley, CA 94704 USA}
\affiliation{Institute for Advanced
    Biosciences, Keio University, 252-8520 Fujisawa, Japan}

\author{Sorin T\u{a}nase-Nicola}
\email[E-mail:]{sorin@amolf.nl}
\altaffiliation[Present address:]{Department of Physics, University of
Michigan, Ann Arbor MI 48109-1040}
\author{Pieter Rein ten Wolde}
\email[E-mail:]{tenwolde@amolf.nl}
\affiliation{FOM Institute for Atomic and Molecular Physics (AMOLF),
Science park 113,
1098 XG Amsterdam, The Netherlands}







\newcommand{\KK}{\rm KK}
\newcommand{\KKi}{\rm KK^*}
\newcommand{\K}{\rm K}
\newcommand{\Kp}{\rm K_p}
\newcommand{\Kpp}{\rm K_{pp}}
\newcommand{\KKtK}{\rm KK-K}
\newcommand{\KKtKp}{\rm KK-K_p}
\newcommand{\Ptase}{\rm P}
\newcommand{\Ptasei}{\rm P^*}
\newcommand{\PtKpp}{\rm P-K_{pp}}
\newcommand{\PtKp}{\rm P-K_{p}}
\newcommand{\cKK}{\rm [KK]}

\begin{abstract}
  Multisite covalent modification of proteins is omnipresent in
  eukaryotic cells.  A well-known example is the mitogen-activated
  protein kinase (MAPK) cascade, where in each layer of the cascade a
  protein is phosphorylated at two sites. It has long been known that
  the response of a MAPK pathway strongly depends on whether the
  enzymes that modify the protein act processively or distributively:
  a distributive mechanism, in which the enzyme molecules have to
  release the substrate molecules in between the modification of the
  two sites, can generate an ultrasensitive response and lead to
  hysteresis and bistability.  We study by Green's Function Reaction
  Dynamics, a stochastic scheme that makes it possible to simulate
  biochemical networks at the particle level and in time and space, a
  dual phosphorylation cycle in which the enzymes act according to a
  distributive mechanism. We find that the response of this network
  can differ dramatically from that predicted by a mean-field analysis
  based on the chemical rate equations. In particular, rapid
  rebindings of the enzyme molecules to the substrate molecules after
  modification of the first site can markedly speed up the response,
  and lead to loss of ultrasensitivity and bistability. In essence,
  rapid enzyme-substrate rebindings can turn a distributive mechanism into a processive
  mechanism.  We argue that slow ADP release by the enzymes can
  protect the system against these rapid rebindings, thus enabling
  ultrasensitivity and bistability.
\end{abstract}

\maketitle

\section{Introduction}
Mitogen-activated-protein kinase (MAPK) cascades are ubiquitous in
eukaryotic cells. They are involved in cell differentiation, cell
proliferation, and apoptosis \cite{Chang:2001le}. MAPK pathways
exhibit very rich dynamics.  It has been predicted mathematically and
shown experimentally that they can generate an ultrasensitive response
\cite{Huang96,Ferrell96,Ferrell:1997lc} and exhibit bistability via
positive feedback \cite{Ferrell:1998vo}.  It has also been predicted
that they can generate oscillations
\cite{Kholodenko:2000oq,Wang:2006dq,Chickarmane:2007eu}, amplify weak but
attenuate strong signals \cite{Locasale:2007bh}, and give rise to
bistability due to enzyme sequestration \cite{Markevich:2004nx,Elf:2004fk}.  MAPK
pathways are indeed important for cell signalling, and for this reason
they have been studied extensively, both theoretically
\cite{Huang96,Ferrell96,Kholodenko:2000oq,Wang:2006dq,Chickarmane:2007eu,Locasale:2007bh,Markevich:2004nx,Elf:2004fk,Levchenko:2000ul,Heinrich:2002pd,Angeli04,Locasale:2008ya,Hornberg:2005ei,Qiao:2007wx,TanaseNicola06,Berezhkovskii:2009jv}
and experimentally
\cite{Huang96,Ferrell:1997lc,Ferrell:1998vo,Wang:2006dq,Hornberg:2005ei,Burack:1997hz,Zhao:2001tt,Santos:2007dp}. However,
in most theoretical analyses, the pathway is modelled using chemical
rate equations
\cite{Huang96,Ferrell96,Kholodenko:2000oq,Chickarmane:2007eu,Markevich:2004nx,Levchenko:2000ul,Heinrich:2002pd,Angeli04,Hornberg:2005ei}. This is a
mean-field description, in which it is assumed that the system is
well-stirred and that fluctuations can be neglected. Here, we perform
particle-based simulations of one layer of the MAPK cascade using our
recently developed Green's Function Reaction Dynamics algorithm
\cite{VanZon05,VanZon05_2}. Our simulations reveal that
spatio-temporal correlations between the enzyme and substrate
molecules, which are ignored in the commonly employed mean-field
analyses, can have a dramatic effect on the nature of the
response. They can not only speed up the response, but
also lead to loss of ultrasensitivity and bistability.

The response time, the sharpness of the input-output relation, and
bistability are key functional characteristics of signal transduction
pathways. The response time does not only determine how fast a cell
can respond to a changing environment, but has also been implicated to
underlie many cellular decisions. For example, processes such as cell
proliferation and differentiation, selection of
T cells, apoptosis, and cell cycle progression are believed to be
regulated by the duration of the signal
\cite{Locasale:2008ya,Marshall:1995zr,Chen:1996ib,Fischle:2001nq,Murphy:2002kc,Murphy:2006ta,Santos:2007dp}. The
sharpness of the input-output relation, or the gain, is a key property
of any signal transduction pathway, since it directly affects the
signal-to-noise ratio.  Bistability can lead to a very
sharp, all-or-none response \cite{Ferrell:1998vo}, buffer the cell
against fluctuations in an input signal, and makes it possible to lock
the cell in a given state. Indeed, bistability, or more in general
multistability, plays a central role in cell differentiation
\cite{Gilbertbook,Manu:2009id}. It is thus important to understand the
mechanisms that underlie bistability, the gain and the response time
of MAPK pathways.

\begin{figure}[t]
\center 
\includegraphics[width=4cm]{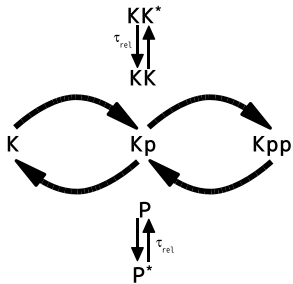}
\caption{Dual phosphorylation cycle of one layer of the MAPK cascade.
  MAPK (K) is activated via double phosphorylation by the kinase MAPKK
  (KK) of the upstream layer and deactivated via dephosphorylation by
  a phosphatase (P). It is assumed that the enzymes KK and P act distributively
  and  become inactive (${\rm KK}^{*}$ and ${\rm
   P}^{*}$) immediately after the substrate has been modified,
 relaxing back to the active state with a characteristic time scale
 $\tau_{\rm rel}$.
\label{fig:mapk_scheme}}
\end{figure}

A MAPK cascade consists of three layers, where in each layer a kinase
activates the kinase of the next layer.  Importantly, full activation
of the kinase requires that it becomes doubly phosphorylated (see
Fig. \ref{fig:mapk_scheme}).
Kinase activation is regulated via a dual
phosphorylation cycle, in which the upstream kinase and a phosphatase
control the phosphorylation state of the two sites of the kinase in an
antagonistic manner. A key question is whether the enzymes that modify
the kinase act in a processive or in a distributive manner
\cite{Huang96,Ferrell96,Ferrell:1997lc}. In a distributive mechanism,
the enzyme has to release the substrate after it has modified the
first site, before it can rebind and modify the second site.  In
contrast, in a processive mechanism, the enzyme remains bound to
 the substrate in between the modification of the two
sites. While a processive mechanism requires only a single
enzyme-substrate encounter for the modification of both sites, a
distributive mechanism requires at least two enzyme-substrate
encounters.

Mean-field analyses based on the chemical rate equations have revealed
that whether the enzymes act according to a processive or a
distributive mechanism has important functional consequences for the
response of a MAPK pathway. A distributive mechanism can generate an
ultrasensitive response since the concentration of the fully activated
kinase depends quadratically on the upstream kinase concentration
\cite{Huang96,Ferrell96,Ferrell:1997lc}. Moreover, if the enzymes are present in limiting
amounts, enzyme sequestration can lead to bistable
behavior if they act distributively \cite{Markevich:2004nx}. 
These mean-field analyses, however, assume that at each instant the molecules
are uniformly distributed in space. Here, we show using
particle-based simulations that spatio-temporal
correlations between the enzyme and the substrate molecules can
strongly affect the response of a MAPK pathway.

We perform particle-based simulations of one layer of a MAPK pathway
in which the enzymes act according to a distributive mechanism. The
simulations reveal that after an enzyme molecule has dissociated from
a substrate molecule upon phosphorylation of the first site, it can
rebind to the same substrate molecule to modify its second site before
another enzyme molecule binds to it. Importantly, the probability per
unit amount of time that such a rebinding event occurs does not depend
upon the enzyme concentration. As a result, enzyme-substrate rebindings can
effectively turn a distributive mechanism into a processive one, even
though modification of both sites of a substrate molecule involves at
least two collisions with an enzyme molecule. Indeed, a distributive
mechanism not only requires a two-collision mechanism, it also
requires that the rates at which they occur depend upon the
concentration.

These rebindings have important functional consequences. Since
rebindings effectively turn a distributive mechanism into a processive
one, ultrasensitivity and bistability via enzyme sequestration are
lost. Moreover, rebindings strongly reduce the gain of the network. We
investigate in depth the scenarios in which rebindings become
important. This reveals that the importance of rebindings depends on
the concentration and the diffusion constant of the molecules: the
lower the concentration and/or the diffusion constant, the more likely
an enzyme molecule rebinds a substrate molecule to modify the second
site before another enzyme molecule does. Since enzyme-substrate
rebindings are faster than random enzyme-substrate encounters,
this observation leads to the counter-intuitive prediction that slower
diffusion can lead to a {\em faster} response. We also find that the
impact of rebindings strongly depends on the time it takes to
re-activate the enzyme after it has modified the first site.  If, for
instance, the ADP/ATP exchange on a kinase has to take place after the kinase
has dissociated from the substrate upon phosphorylation of the first
site, but before it can bind the substrate again to modify the second
site, then either slow ADP release or slow ATP supply will make
enzyme-substrate rebindings less important. ADP release from protein
kinases has been reported to be fairly slow \cite{Keshwani:2008tr},
suggesting that slow ADP release might
be critical for generating ultrasensitivity and bistability.

The importance of rebindings relies on the interplay between reaction
and diffusion at short and long length and time scales. This means
that the algorithm should correctly capture the spatio-temporal
dynamics of the system at both scales. In this manuscript, we present
and apply an enhanced version of our recently developed Green's
Function Reaction Dynamics algorithm. This particle-based algorithm is not only
even more efficient than the original GFRD scheme, which is already 4
to 5 orders more efficient than brute-force Brownian Dynamics
\cite{VanZon05}, it is also exact. 

Biological systems that exhibit
macroscopic concentration gradients or spatio-temporal oscillations,
which have recently been studied extensively, are typically considered
to be reaction-diffusion problems. We believe that our simulations are
the first to show that in a biological system that is spatially
uniform, spatio-temporal correlations on molecular length scales can
drastically change the {\em macroscopic} behaviour of the system. This
underscores the importance of particle-based modelling of biological
systems in time and space.

\section{Model}
\subsection{Dual phosphorylation cycle}
We consider one layer of the MAPK pathway, consisting of one dual
modification cycle, as shown in Fig. \ref{fig:mapk_scheme}.
Phosphorylation and dephosphorylation proceed via Michaelis-Menten
kinetics and according to an ordered, distributive mechanism. Importantly, we
assume that the enzymes are inactive after they have released their
modified substrate; before they can catalyse the next reaction, they
first have to relax back to the active state. The inactive state could
reflect that the enzyme is in an inactive conformational state after
it has released its product. For the kinase it could also reflect that
after it has released its substrate, ADP is bound; only when ADP has
been released and ATP has been bound, does the enzyme become active
again. As we will discuss in detail below, the timescale for
re-activation, $\tau_{\rm rel}$, plays a key role in the dynamics of the system.

This model is described by the following reactions:
\begin{eqnarray}
\KK + \K \overset{k_{1}}{\underset{k_{2}}{\rightleftharpoons}}&
\KKtK\stackrel{k_{3}}{\rightarrow}&\KKi + \Kp\label{eq:K1}\\
\KK + \Kp  \overset{k_4}{\underset{k_5}{\rightleftharpoons}} &\KKtKp
\stackrel{k_6}{\rightarrow}&\KKi + \Kpp \label{eq:K2}\\
\Ptase + \Kpp \overset{k_{1}}{\underset{k_{2}}{\rightleftharpoons}}&
\PtKpp\stackrel{k_{3}}{\rightarrow}&\Ptasei + \Kp\label{eq:K3}\\
\Ptase + \Kp \overset{k_4}{\underset{k_5}{\rightleftharpoons}}&
\PtKp\stackrel{k_6}{\rightarrow}&\Ptasei + \K\label{eq:K4}\\
\KKi\stackrel{k_{7}}{\rightarrow}&\KK, \ \ \ \ \
\Ptasei\stackrel{k_{7}}{\rightarrow}&\Ptase\label{eq:K5}
\end{eqnarray}
The first two reactions describe the phosphorylation of the kinase of
interest, MAPK (K), by the upstream kinase, MAPKK (KK), while
Eqs. \ref{eq:K3} and \ref{eq:K4} describe its dephosphorylation by the
phosphatase (P). The inactive state of the enzymes after they have
released their product is denoted by the superscript $^*$, and the
relaxation towards the active state is described by the last two
equations. For simplicity, we assume that re-activation can be
described as a simple unimolecular reaction with a time scale
$\tau_{\rm rel} \simeq 1 / k_7$. We also assume that the system is
symmetric, meaning that the rate constants for the phosphorylation
reactions are equal to the corresponding rate constants for the
dephosphorylation reactions. We will systematically vary the
relaxation time $\tau_{\rm rel}$, and the concentration and the diffusion
constant of the particles, $D$ (see below). For the other parameter
values, we have taken typical values from the literature (see
Methods).

\subsection{Green's Function Reaction Dynamics}
  We will compare the predictions of a mean-field model based on the
  chemical rate equations \cite{Markevich:2004nx} with those of a
  model in which the particles are explicitly described in time and
  space. In this particle-based model, it is assumed that the
  molecules are spherical in shape, have a diameter $\sigma$, and move
  by diffusion with a diffusion constant $D$. Moreover, two reaction
  partners can react with each other with an intrinsic rate
  $k_{\rm a}=k_1$ or $k_4$, respectively, once they are in contact,
  and two associated species can dissociate with an intrinsic
  dissociation rate $k_{\rm d}=k_2$ or $k_5$, respectively.

One algorithm to simulate this particle-based model would be Brownian Dynamics. However, since the concentrations are fairly low,
much CPU time would be wasted on propagating the reactants towards
one another. We therefore employ our recently developed Green's
Function Reaction Dynamics algorithm, which uses Green's functions to
concatenate the propagation of the particles in space with the
chemical reactions between them, allowing for an event-driven
algorithm \cite{VanZon05,VanZon05_2} (see Methods).

\section{Results}
\subsection{Rebindings}
\begin{figure}[t]
\includegraphics[width=6.5cm]{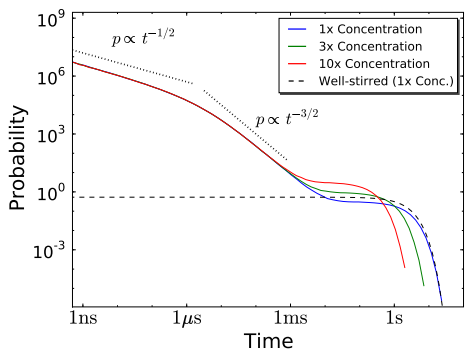}
\caption{The distribution of association times for a bimolecular
  reaction for different concentrations.  The system consists of one A
  molecule that can associate with $N=10$ B molecules according to the
  reaction $A + B
  \overset{k_{a}}{\underset{k_{d}}{\rightleftharpoons}} C$.  For $t <
  \tau_{\rm mol} \approx \sigma^2/D \approx 1\mu{\rm s}$ the
  distribution decays as $t^{-1/2}$, for $\tau_{\rm mol} < t < \tau_{\rm
    bulk} \approx 1{\rm s}$ it decays as $t^{-3/2}$, while for
  $t>\tau_{\rm bulk}$ the distribution decays exponentially. The
  algebraic decay is due to rebinding events, in which a dissociated B
  molecule rebinds the A molecule before diffusing into the bulk; this
is   unaffected by the concentration of B, $[{\rm B}]$.  The exponential
  relaxation is due to B molecules that arrive at A from the bulk, and
  is a function of $[{\rm B}]$. The concentration was controlled by changing
  the volume from $V = 1 $, $0.33$, and $0.1\, {\rm \mu m^3}$,
  corresponding to ${\rm [B]} = N / V = 16$, $48$, and $160\,{\rm
    nM}$, respectively.  $k_{\rm a} = 0.056\, {\rm nM}^{-1}{\rm
    s}^{-1}$, $k_{\rm d} = 1.73\, {\rm s^{-1}}$, $D = 1 \mu{\rm m}^2
  {\rm s^{-1}}$ and $\sigma = 5\, {\rm
    nm}$.
\label{fig:rebindings}}
\end{figure}

To understand the response of the dual phosphorylation cycle, it is
critical to consider the distribution of association times for a
bimolecular reaction. We consider a simple
bimolecular reaction, $A+B \rightleftharpoons C$, where one A molecule
can react with one of $N$ B molecules to form a C molecule in a
volume $V$. A model in which it is assumed that the particles are
uniformly distributed in space at all times, be it a mean-field
continuum or a
stochastic discrete model, predicts that this distribution is exponential (see
Fig. \ref{fig:rebindings}).
In contrast, in a spatially-resolved
model, the distribution of association times is algebraic on short
times and exponential only at later times \cite{VanZon06}.

The difference between the well-stirred model and the
spatially-resolved model is due to rebindings. In a well-stirred
model, the propensity that after a dissociation event the A molecule
reacts with a B molecule only depends on the total density $N/V$ of B
molecules, and not on their positions---in a spatially resolved model
this would amount to putting the dissociated B particle to a random
position in the cell. Since the total density of B is constant, the
association propensity is constant in time, leading to an exponential
waiting-time distribution in the well-stirred model. In the spatially
resolved model the situation is markedly different. The B molecule
that has just dissociated from the A molecule is in close proximity to
the A molecule.  As a consequence, it can rapidly rebind to the A
molecule before it diffuses away from it into the bulk. Such
rebindings lead to the algebraic decay of the association-time
distribution at short times. For times shorter than the time to travel
a molecular diameter, $ t<\tau_{\rm mol} \approx \sigma^2/D$ (see {\em
  Supporting Information}), the
dissociated B particle essentially experiences a surface of the A
particle that is flat, and its rebinding dynamics is given by that of
a 1D random walker returning to the origin, leading to the $t^{-1/2}$
decay. At times $\tau_{\rm mol} < t < \tau_{\rm bulk}$, the
dissociated B particle sees the entire sphere of A, and the
probability of a re-encounter event is that of a 3D random walker
returning to the origin, decaying as $t^{-3/2}$. At times $t >
\tau_{\rm bulk}$, the dissociated B particle has diffused into the
bulk, and it has lost all memory where it came from.  The probability
that this molecule, or more likely, another B molecule binds the A
molecule, now becomes constant in time, leading to an exponential
waiting-time distribution at long times \cite{VanZon06}.  

Fig. \ref{fig:rebindings} 
shows that the association-time distribution
depends on the concentration for $t>\tau_{\rm bulk}$, but not for $t <
\tau_{\rm bulk}$. Indeed, while the encounter rate between two
molecules in the bulk depends on their concentration, the rate at which
a rebinding event occurs is independent of it.  As we will show below, this
has major functional consequences for the response of the dual
phosphorylation cycle.

\subsection{Rebindings can speed up the response}
\begin{figure}[t]
\includegraphics[width=6.5cm]{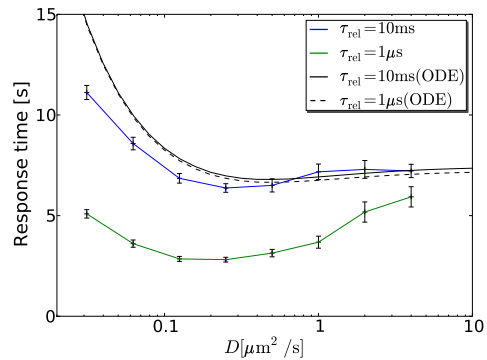}
\caption{Average response time as a function of the diffusion constant $D$
  for $\tau_{\rm rel} = 10\, {\rm ms}$ (blue line) $\tau_{\rm rel} =
  1\, \mu{\rm s}$ (green line), as predicted by the particle-based
  model; for comparison, the predictions of the mean-field model based
  on the ODE chemical rate equations are also shown (black
  lines). Initially, only the phosphatases are active; at $t=0$ the
  upstream kinases are activated, and plotted is the time it takes on
  average to reach 50\% of the final steady-state level of doubly
  phosphorylated substrate (Kpp). The optimum in the particle-based
  model is due to the interplay between phosphorylation of the first
  site, which slows down with decreasing diffusion constant since
  enzyme and substrate have to find each other at random, and
  phosphorylation of the second site, which speeds up with decreasing
  diffusion constant, because of enzyme-substrate
  rebindings.
  \label{fig:response_time}}
\end{figure}

Fig. \ref{fig:response_time} 
shows the average response time as a function
of the diffusion constant, for two different values of the lifetime of
the inactive state of the enzymes, $\tau_{\rm rel}$. The figure
reveals that both the mean-field (ODE) and the particle-based model
predict that there is an optimal diffusion constant that minimizes the
response time. However, in the mean-field model the optimum is barely
noticeable \cite{Footnote1}\cite{Agmon90}.  To a good approximation,
the mean-field model predicts that the response time increases with
decreasing diffusion constant, because enzyme-substrate association
slows down as diffusion becomes slower. In contrast, the
particle-based model shows a marked optimum, which is most pronounced
when $\tau_{\rm rel}$ is short. Clearly, the particle-based
simulations predict that slower diffusion can lead to a faster
response.

\begin{figure*}[t]
(A) \includegraphics[width=5cm]{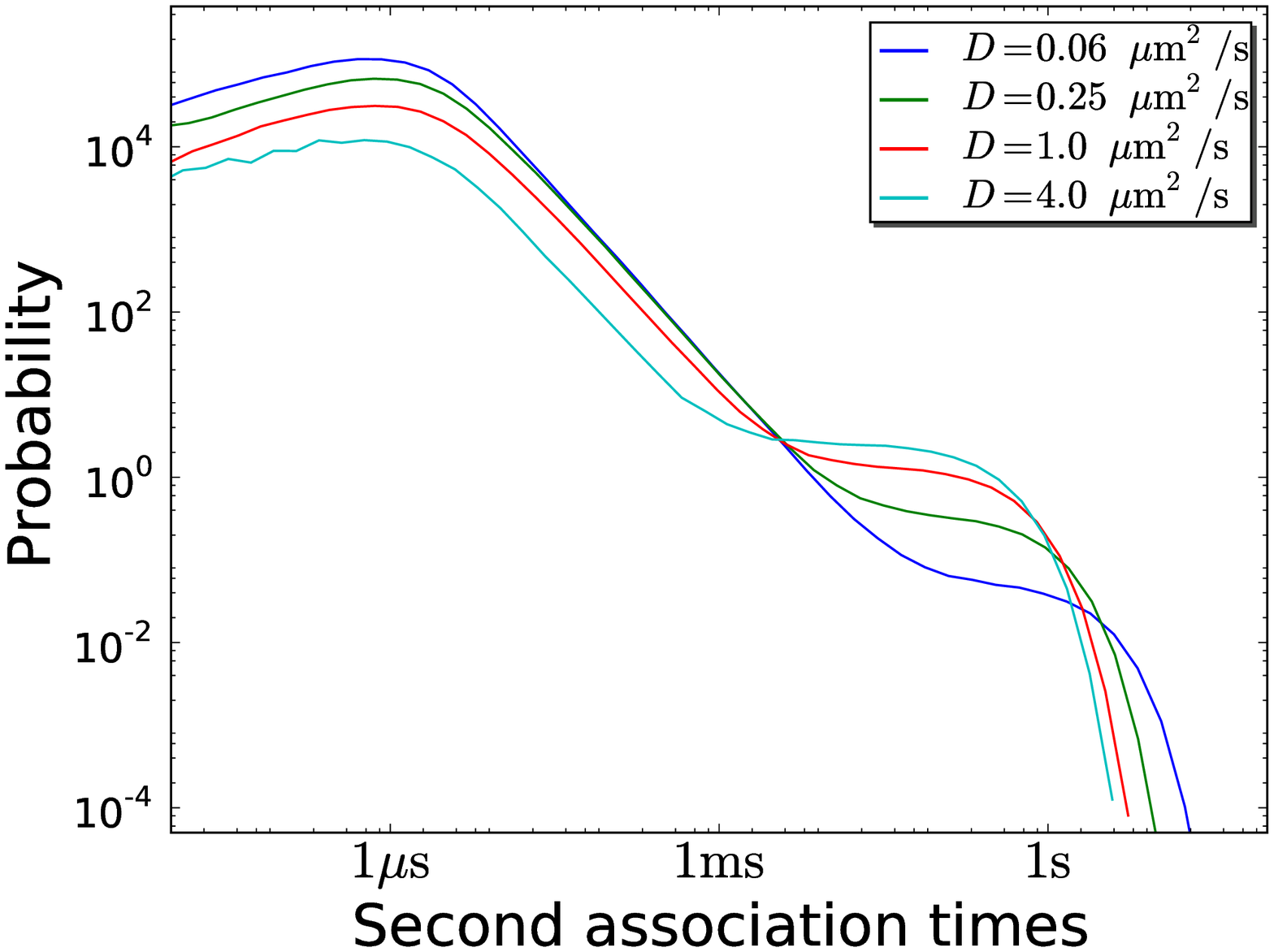}
(B) \includegraphics[width=5cm]{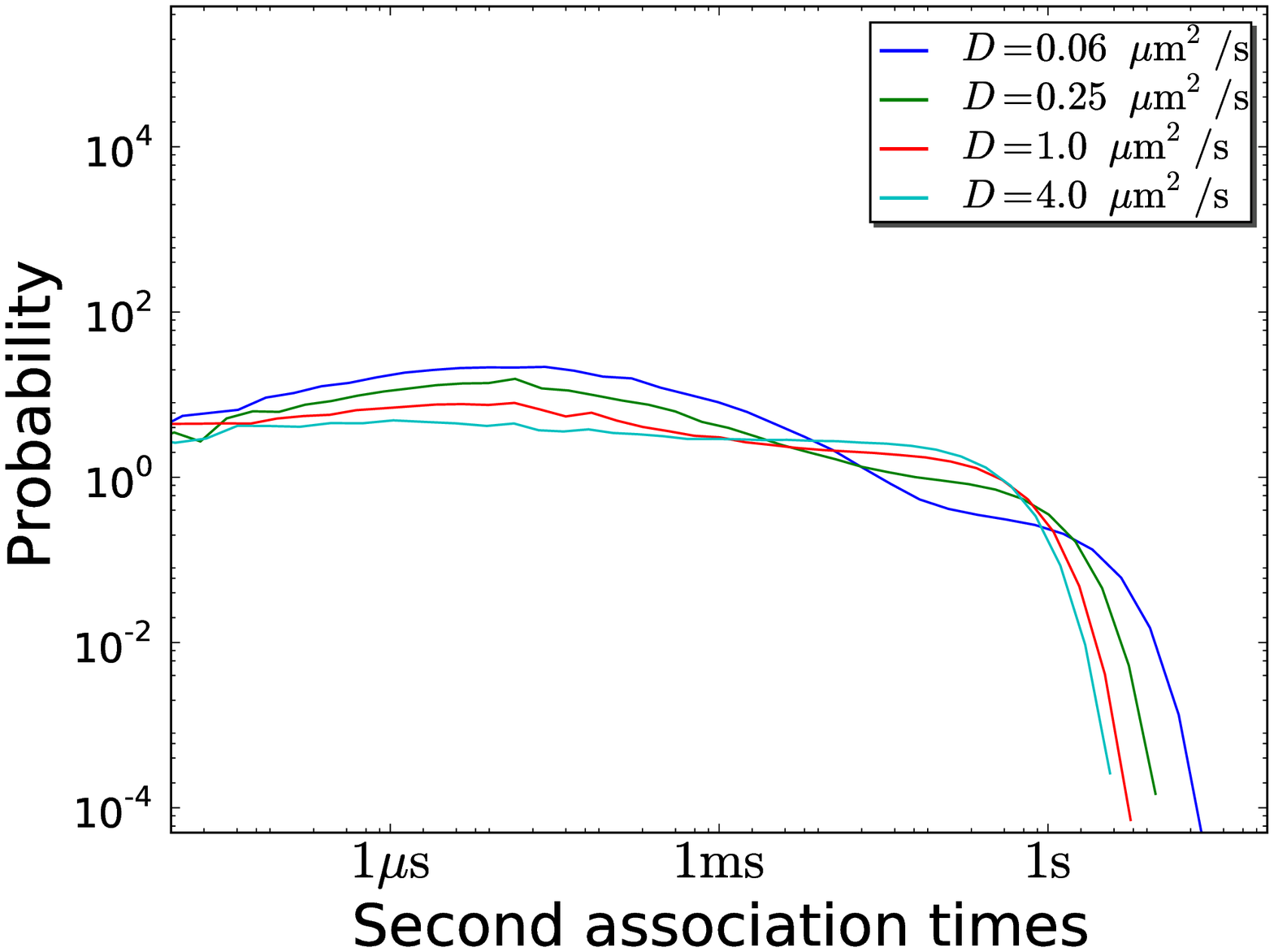}
(C) \includegraphics[width=5cm]{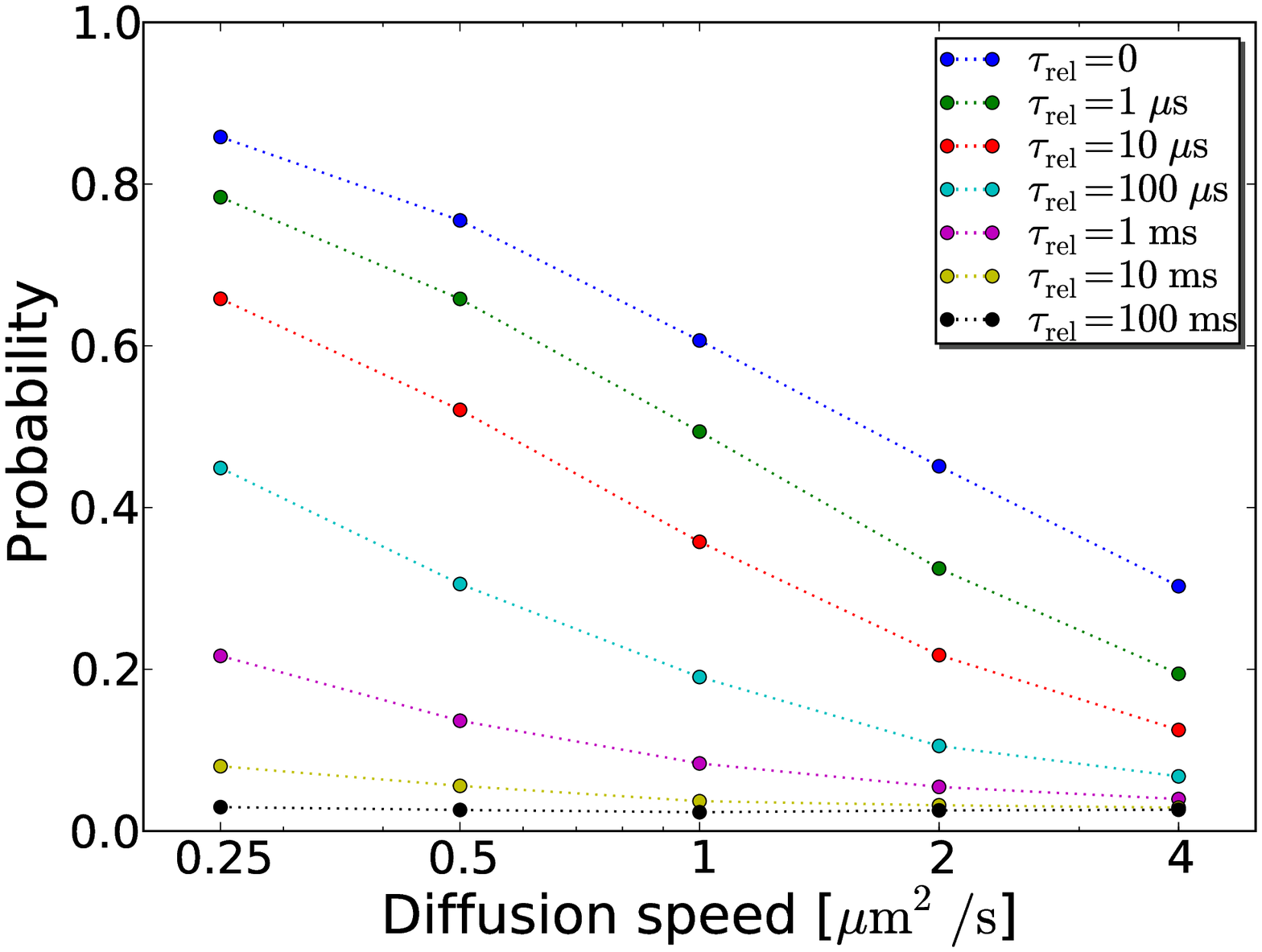}
\caption{Distribution of times it takes for a substrate that has just
  been phosphorylated once (Kp) to bind a kinase molecule (KK), for
  different diffusion constants. The enzyme reactivation time is (A)
  $\tau_{\rm rel} = 1\mu{\rm s}$ or (B) $\tau_{\rm rel} = 10 {\rm
    ms}$; in both cases $\tau_{\rm mol} \approx 1 \mu{\rm s}$ and
  $\tau_{\rm bulk} \approx 0.1 - 10 {\rm s}$. (C) Probability that the two modification sites of a
  substrate molecule are phosphorylated by the
  same kinase molecule as a function of diffusion constant, for
  different enzyme re-activation times $\tau_{\rm rel}$. It is seen
  that the probability of an enzyme-substrate rebinding event
  increases not only with decreasing enzyme reactivation time, but
  also with decreasing diffusion constant. The latter explains why
  slower diffusion can lead to a faster response, as seen in Fig. 
  \ref{fig:response_time}.\label{fig:P_tp2}} 
\end{figure*}

The speed up of the response with slower diffusion is due to the
interplay between enzyme-substrate rebindings, and enzyme
re-activation.  This interplay manifests itself in the distribution of
the {\em second-association time}, defined as the time it takes for a
substrate molecule that has just been phosphorylated (Kp) to bind a
kinase molecule (KK) for the phosphorylation of the second site
(Fig. \ref{fig:P_tp2}).
After a kinase molecule (KK) has
phosphorylated the first site of a substrate molecule (K), it will
dissociate from it. After dissociation, it is still in close proximity
to the substrate molecule, and it will therefore rapidly re-encounter
the substrate molecule before it diffuses away into the bulk. When the
lifetime $\tau_{\rm rel}$ of the inactive state of the kinase molecule
is short compared to the time $\tau_{\rm mol}$ it takes for the enzyme and substrate
molecule to diffuse away from each other, the probability that upon a re-encounter the enzyme molecule
has become active again such that it can actually rebind the substrate
molecule, will be large. Hence, when $\tau_{\rm rel} \leq \tau_{\rm
  mol}$, the kinase will often rapidly rebind the substrate molecule,
leading to the characteristic algebraic decay of $t^{-3/2}$ for $
\tau_{\rm mol} < t < \tau_{\rm bulk}$
(Fig. \ref{fig:P_tp2}A).
However, there is also a probability that the
enzyme molecule will escape into the bulk before it rebinds the
substrate molecule. If this happens, most likely another kinase
molecule binds the substrate molecule. This scenario underlies the
exponential form of the second-association-time distribution at longer
times, with the corner time $\tau_{\rm bulk} \approx 0.1-10{\rm s}$. It can now also be
understood why the marked peak in the distribution at short times
(Fig. \ref{fig:P_tp2}A)
disappears when the enzymes' reactivation time
$\tau_{\rm rel}$ becomes significantly longer than $\tau_{\rm mol}$
(Fig. \ref{fig:P_tp2}B):
after phosphorylation of the first site, the
kinase will rapidly re-encounter the substrate molecule many times,
but since the enzyme is most probably still inactive, it cannot rebind
the substrate molecule, and it will therefore diffuse into the
bulk. In the {\em Supporting Information} we derive analytical
expressions for the enzyme-substrate rebinding-time distributions, and
elucidate the different scaling regimes that can be observed.

To understand why slower diffusion can lead to a faster response when
the lifetime of the enzymes' inactive state is short
(Fig. \ref{fig:response_time}),
it is instructive to consider how the
distribution of second-association times depends on the diffusion
constant. Fig. \ref{fig:P_tp2}A 
shows that the corner at $
\tau_{\rm bulk}$ shifts to longer times as
the diffusion constant is decreased. This is because the rate at which
a kinase molecule from the bulk encounters a given substrate molecule
is given by $1/\tau_{\rm bulk} = k_{\rm D} = 4\pi \sigma (D_{\rm E} +
D_{\rm S}) \cKK$, where $\sigma$ is the sum of the
radii of the enzyme and substrate molecules and $D_{\rm E}$ and
$D_{\rm S}$ are the diffusion constants of the enzyme and substrate
molecules, respectively. Clearly, substrate phosphorylation by kinase
molecules that have to find the substrate molecules at random slows
down as the molecules move slower.  However, the figure also shows
that the distribution at the corner time of $\tau_{\rm bulk}$
decreases in magnitude while the peak at $\tau_{\rm mol}$ increases in
magnitude when diffusion becomes slower. This means that as the
diffusion constant becomes lower, phosphorylation of the second site
is increasingly dominated by enzyme-substrate rebindings rather than
by random enzyme-substrate encounters. The probability that the enzyme molecule is still in the
vicinity of the substrate molecule after it has relaxed back to the
active state, increases as the diffusion constant decreases, making a
substrate-rebinding event more likely. This is demonstrated
quantitatively in Fig. \ref{fig:P_tp2}C,
which shows the probability
that both sites on the substrate are phosphorylated by the same kinase
molecule. As expected, this probability not only increases with
decreasing lifetime of the enzymes' inactive state, but also with
decreasing diffusion constant. Since enzyme-substrate rebindings are
more rapid than random enzyme-substrate encounters, this
explains why slower
diffusion can lead to a faster response.

While slower diffusion speeds up the modification of the second
site by making rapid enzyme-substrate rebindings more likely,
it also slows down the modification rate of the first site
since that is determined by the rate at which enzyme molecules find
the substrate molecules from the bulk. This is the origin of the
optimum diffusion constant that minimizes the response time
(Fig. \ref{fig:response_time}).

\subsection{Enzyme-substrate rebindings can weaken the sharpness of the response}
\begin{figure}[t]
\hspace*{-9mm} (A) \hspace*{-2mm} \includegraphics[width=4.1cm]{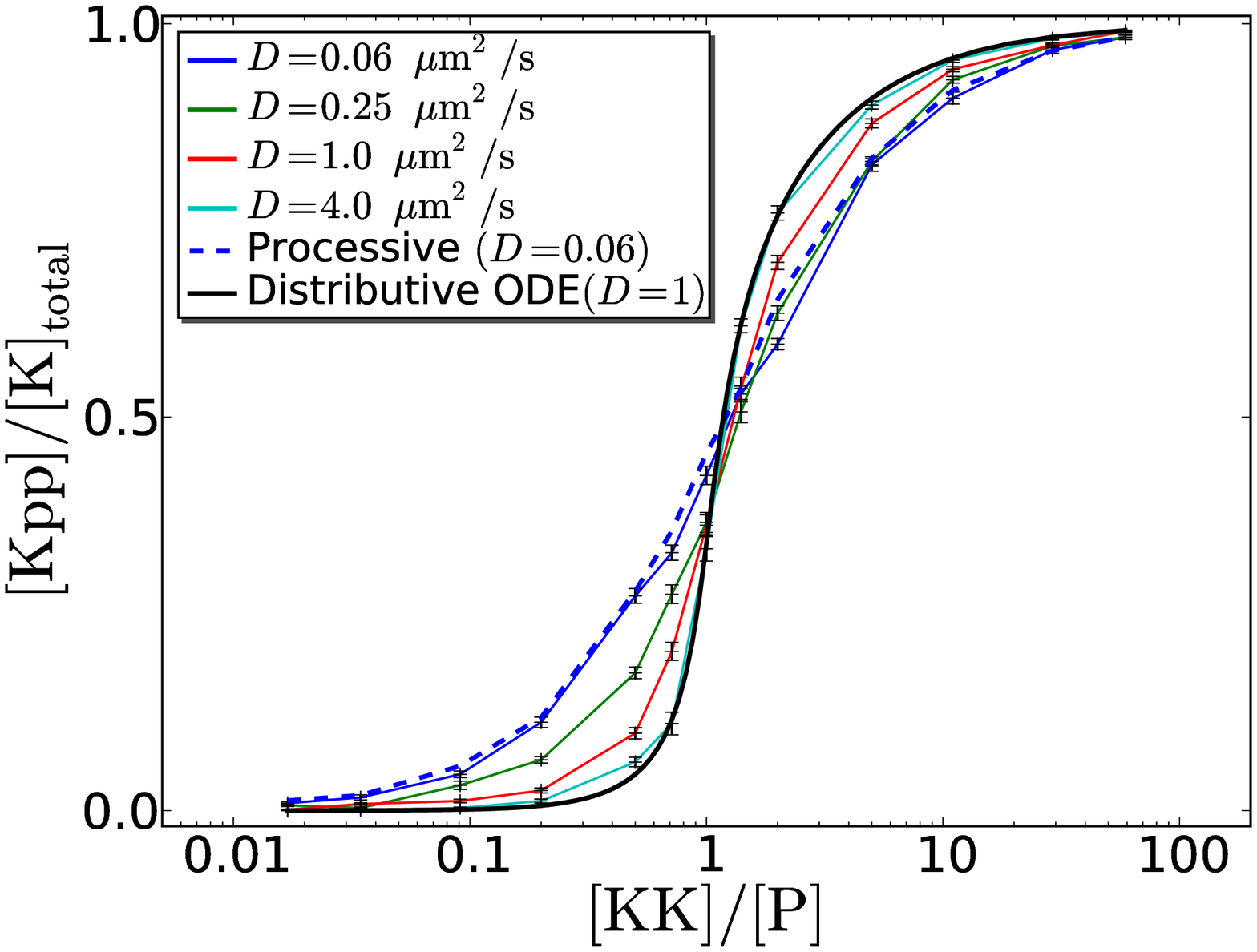}
(B) \hspace*{-2mm} \includegraphics[width=4.1cm]{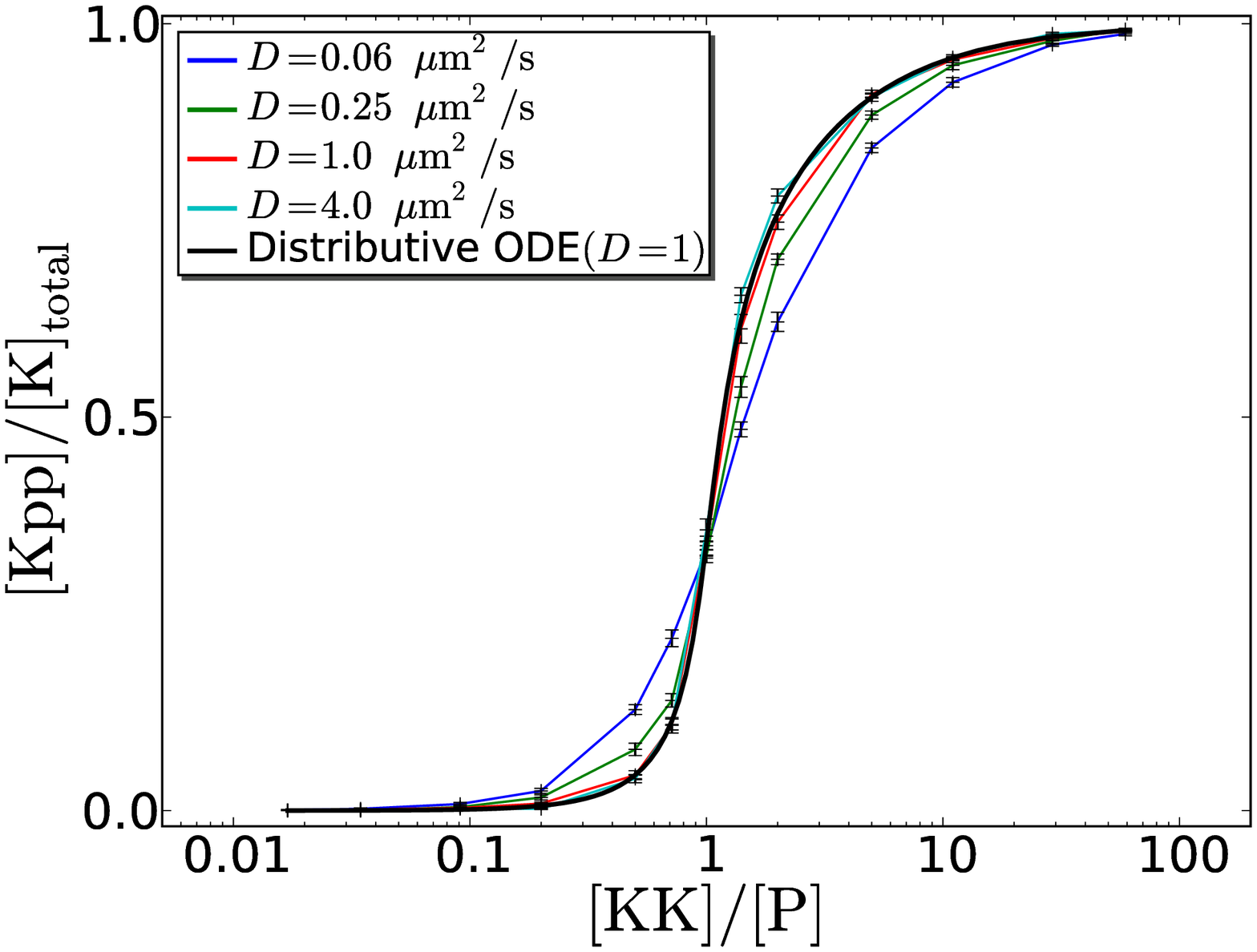}
\caption{Steady-state input-output relations for different diffusion
  constants when (A) $\tau_{\rm rel} = 1\, {\rm \mu s}$ and (B)
  $\tau_{\rm rel} = 10\, {\rm ms}$.  For comparison, we also show the
  predictions of a mean-field (ODE) model of a distributive system with
    $D = 1\, {\rm \mu m^2 / s}$, and that of a particle-based model
    of a {\em processive} system with  $D = 0.06\, {\rm \mu m^2 / s}$
    (only in panel A).
  Note that when the re-activation time $\tau_{\rm rel}$ is short,
  the input-output relation of a distributive system approaches that
  of a processive one as the diffusion constant is lowered (panel A;
  blue lines).\label{fig:in_out}}
\end{figure}

Fig. \ref{fig:in_out} 
shows the effect of enzyme-substrate rebindings
on the steady-state input-output relation. It is seen that when the
re-activation time of the enzymes is long, $\tau_{\rm rel} = 10\, {\rm
  ms}$, the input-output relation is strongly sigmoidal
(Fig. \ref{fig:in_out}B).
Moreover, it does not depend much on the
diffusion constant of the molecules, and it agrees quite well with
that predicted by the mean-field model based on the chemical rate
equations (Fig. \ref{fig:in_out}B
In contrast, when $\tau_{\rm rel}$
is short, {\em i.e.} $\tau_{\rm rel} = 1\mu{\rm s}$, the input-output
relation markedly depends on the diffusion constant
(Fig. \ref{fig:in_out}A).
For large diffusion constants, the response
curve agrees well with that predicted by the mean-field model of a
distributive mechanism. But for lower diffusion constants, it
increasingly deviates from the mean-field prediction, and it becomes
significantly less sigmoidal.

It is commonly believed that multi-site covalent modification can lead
to a sigmoidal, cooperative response when the enzymes act
distributively, but not when they act processively
\cite{Huang96,Gunawardena:2005bu}.  While in a distributive scheme
modification of $n$ sites of a substrate molecule requires at least
$n$ enzyme-substrate binding events, in a processive scheme only one
enzyme-substrate binding event is needed.  This is often presented as
the explanation for why a distributive mechanism enhances the
sensitivity of the modification level to changes in enzyme
concentration. However, Fig. \ref{fig:in_out}A 
shows that when the
enzymes' re-activation time is short and the species' diffusion
constant is low, the input-output relation of a distributive, dual
phosphorylation cycle approaches that of a processive, dual
phosphorylation cycle. This is due to enzyme-substrate
rebindings. Even though during a rebinding trajectory the enzyme
molecule is detached from the substrate molecule and two binding
events are required for full substrate modification, the rate at which
the second site is modified does not depend on the enzyme
concentration (Fig. \ref{fig:rebindings}).
The sharpness of the response increases with the number
of required enzyme-substrate binding events, but only when these
depend on the enzyme concentration.  Enzyme-substrate rebindings effectively
turn a distributive mechanism into a processive mechanism.

\subsection{Rebindings can lead to loss of bistability}
\begin{figure}[t]
\includegraphics[width=6.5cm]{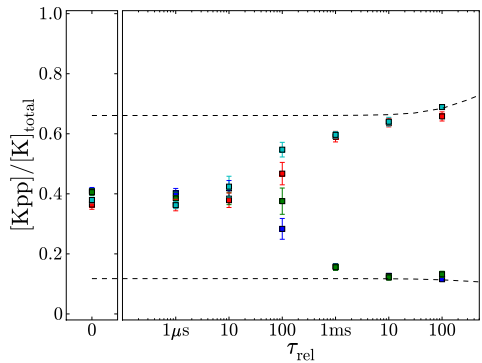}
\caption{Relative Kpp concentrations as a function of the lifetime of
  the inactive state of the enzyme, $\tau_{\rm rel}$.  ${\rm
    [K]_{total} = [Kpp] + [Kp] + [K]}$ was increased to $500 \, {\rm nM}$
  to bring the system to a regime where the mean-field model based on
  the chemical rate equations predicts bistability.  At each value
  of $\tau_{\rm rel}$, the particle-based model was simulated until it
  reaches steady state, starting from four different initial
  conditions, ${\rm [Kpp]/ [K]_{total}} = 0$ (blue), $0.3$ (green),
  $0.7$ (red), and $1$ (cyan).  While the mean-field model shows
  bistability over the whole range of $\tau_{\rm rel}$ (black dotted lines), the
  particle-based model exhibits a bifurcation from mono- to bistability at
  $t \approx 100 \mu{\rm s}$. At this bifurcation point, the system
  critically slows down, as a result of which it does not
  even equilibrate after $350 \, {\rm s}$. \label{fig:bistability}}
\end{figure}

Markevich {\em et al.} have shown that bistability can arise in a dual
phosphorylation cycle when the enzymes act distributively and are
present in limiting concentrations \cite{Markevich:2004nx}.  The idea
is that if the substrate molecules are, for example, predominantly
unphosphorylated and a substrate molecule is phosphorylated to become
singly phosphorylated, it will most likely bind a phosphatase molecule
to become unphosphorylated again, instead of a kinase molecule to
become fully phosphorylated---when most of the substrate molecules are
unphosphorylated, the kinase molecules are mostly sequestered by the
unphosphorylated substrate molecules, while the phosphatase molecules
are predominantly unbound. However, this is essentially a mean-field
argument, which assumes that the substrate and enzyme molecules are
randomly distributed in space at all times. Fig. \ref{fig:bistability}
shows that spatio-temporal correlations between the enzyme and
substrate molecules can have a dramatic effect on the existence of
bistability. When the enzymes' reactivation time $\tau_{\rm rel}$ is long,
spatio-temporal correlations are not important, and the system indeed
exhibits bistability. But when $\tau_{\rm
  rel}$ is short, the probability that a substrate molecule that has
just been phosphorylated once will be phosphorylated twice is larger
than that it will be dephosphorylated again: the chance that it
will rebind the kinase molecule that has just phosphorylated it, will,
because of the close proximity of that kinase molecule, be larger than
the probability that it will bind a phosphatase molecule, even though
in this state there are many more phosphatase than kinase molecules to
which the substrate molecule could bind to.  These rebindings, or more
precisely, spatio-temporal correlations between the enzyme and
substrate
molecules, are the origin of the loss of bistability
when $\tau_{\rm rel}$ is short (Fig. \ref{fig:bistability}).

\subsection{The effect of concentration}
Figs. \ref{fig:response_time}-\ref{fig:bistability}
show that
enzyme-substrate rebindings are significant when the concentration of
enzyme and substrate is on the order of 100 nM, which is a
biologically relevant range \cite{Huang96,Ferrell96,Markevich:2004nx}.
Fig. S4 of the {\em Supporting Information} shows that  when the
concentrations of all species are increased by more than a factor 10
from those used in Fig. \ref{fig:in_out},
the system becomes
bistable. While the distribution of rebinding times does not depend on
the concentration, the competition between phosphatase and kinase
molecules in the bulk for binding to the substrate does, in such a way
that the system is driven deeper into the bistable regime (see Fig. S5
in {\em Supporting Information}). Increasing the concentration can thus
overcome the effect of enzyme-substrate rebindings.

\section{Discussion}
Multi-site phosphorylation is omnipresent in biological
systems. Perhaps the best known and arguably the most studied example
is the dual phosphorylation cycle of the MAPK pathway, studied here,
but other well-known examples are the Kai system \cite{Zon:2007ly},
the CDK inhibitor Sic1 \cite{Nash:2001qs}, the NFAT system
\cite{Crabtree:2002gv}, and the CAMKII system
\cite{Miller:2005ud}. Multi-site phosphorylation can lead to an ultrasensitive response \cite{Huang96,Ferrell96}, to a
threshold response \cite{Gunawardena:2005bu}, to bistability
\cite{Markevich:2004nx,Miller:2005ud}, or synchronise oscillations of phosphorylation levels of individual
protein molecules \cite{Zon:2007ly}, provided the enzymes act via a
distributive mechanism.  We have
studied using a particle-based model a dual phosphorylation cycle in
which the enzymes act according to a distributive mechanism. Our
results show that rapid enzyme-substrate rebindings can effectively
turn a distributive mechanism into a processive mechanism, leading to
loss of ultrasensitivity and bistability. Moreover, our results reveal
that enzyme-substrate rebindings can significantly speed up the
response, with slower diffusion leading to a faster response. While
rebindings have been predicted to affect the noise in signal detection
\cite{VanZon06,Andrews05}, our results predict that they can also
drastically change the macroscopic behaviour of the system.

Our results reveal that enzyme-substrate rebindings occur on short
length and time scales. Rebindings are important up to time scales of
about $1 - 10 \, {\rm ms}$ (Fig. \ref{fig:P_tp2}),
corresponding to
the time for a protein to diffuse over a few molecular
diameters. Beyond those length and time scales the dissociated enzyme
and substrate molecules have essentially lost memory where they came
from, and they would have to find each other again at random. An
important question is whether we should not have taken orientational
diffusion into account, precisely because rebindings occur at
comparable length and time scales. However, the first and second
phosphorylation site are often close to each other on the substrate,
{\em e.g.} separated by only a single amino-acid residue
\cite{Payne:1991up}, suggesting that enzyme-substrate rebindings can
indeed occur without significant orientational diffusion. Moreover,
our model does not include molecular crowding, and it seems likely
that subdiffusion caused by crowding can significantly extend the time
scale over which rebindings occur \cite{Lomholt:2007tg}.

The importance of enzyme-substrate rebindings depends on the lifetime
of the inactive state of the enzymes. For a typical protein diffusion
constant of $D=1-10\,\mu{\rm m}^2{\rm s}^{-1}$
\cite{Elowitz99,Meacci:2006ls,Elf:2007qe}, the rebinding probability
drops below $10 \, \%$ when the enzyme re-activation time becomes
longer than $10\,{\rm ms}$ (Fig. \ref{fig:P_tp2}C).
Slow enzyme
re-activation may thus be critical for generating bistability and
ultrasensitivity. To our knowledge, re-activation times of enzymes in
MAPK pathways have not been measured yet. The re-activation time of a
kinase will depend sensitively on the order in which ADP and modified
substrate dissociate from it, and ATP and substrate bind to it.  If a
kinase can bind its substrate irrespective of the nucleotide binding
state, then nucleotide exchange will not be rate limiting. If the ADP
is released before the modified substrate, but ATP binding is required
for binding of the next substrate, then ATP binding might be the
rate-limiting step; with mM ATP concentrations, this is however
expected to yield fast re-activation times, of order microseconds. If
the modified substrate must dissociate before ADP can dissociate and
ADP must dissociate before the kinase can bind substrate again, then
the rate of ADP release may become rate limiting. A recent study on a
protein kinase provides support for the latter scenario, with an ADP
release rate that is on the order of $100 {\rm ms}$
\cite{Keshwani:2008tr}. This suggests that slow ADP release may allow
for ultrasensitivity and bistability, although more work is needed to
explore these mechanisms in depth.  Concerning bistability, it is
possible that bistability requires the phosphatase to act
distributively \cite{Markevich:2004nx}. Bistability could thus be lost
if the mechanism by which the phosphatase acts changes from a
distributive to a processive mechanism due to rebindings. To our
knowledge, it is unknown what the minimum time is to re-activate a
phosphatase. It is conceivable that this time is very short. Rapid
re-activation of the phosphatase could thus lead to loss of
bistability.

Our results show that experiments to determine whether an enzyme acts
distributively or processively should be interpreted with care. These
experiments are often performed by investigating the time courses of
the concentrations of the intermediate and final products
\cite{Ferrell:1997lc}. If the amount
of intermediate products exceeds that of the enzyme, then the
mechanism must be distributive. However, our results
reveal that the converse does not necessarily imply that the mechanism
is processive, as commonly assumed: enzyme-substrate rebindings can turn a distributive
mechanism into a processive one, with the concentration of the
intermediate product remaining below that of the enzyme. We stress
that the question whether an enzyme acts processively because of
rebindings or because it remains physically attached to the substrate
is biologically relevant, because the importance of enzyme-substrate
rebindings strongly depends on the conditions. It depends on the
diffusion constants of the components, the lifetime of the inactive
state of the enzyme, and on the concentrations of the components. All
these factors may vary from one place in the cell to another and will
vary from one cell to the next. In fact, an enzyme that operates
according to a distributive mechanism in the test-tube may act
processively in the crowded environment of the cell.

Finally, how could our predictions be tested experimentally? If the
enzyme of interest is a kinase, then one experiment would be to change
the lifetime of the inactive state by varying the ATP concentration or
by making mutations that change the ADP release rate.  Another
proposal would be to study the enzyme kinetics as a function of the
concentration of a crowding agent, such as PEG
\cite{Keshwani:2008tr}. Crowding will slow down diffusion, and will,
because of subdiffusion \cite{Lomholt:2007tg}, increase the time that an enzyme and a
substrate molecule that are in close proximity, stay together. Both
effects will make enzyme-substrate rebindings more likely. Studying
the input-output relation and the time course of the intermediate and
final products \cite{Ferrell:1997lc,Keshwani:2008tr} for different
levels of macromolecular crowding will shed light on the importance of
spatio-temporal correlations for the macroscopic behavior of
biological systems employing multi-site modifications.

\section{Methods}
\subsection{Green's Function Reaction Dynamics}
A reaction-diffusion system is a many-body problem that can not be
solved analytically. The key idea of GFRD is to decompose the
many-body problem into single and two-body problems, which can be
solved analytically using Green's functions
\cite{VanZon05,VanZon05_2}. These Green's functions are then used to
set up an event-driven algorithm, which makes it possible to make
large jumps in time and space when the particles are far apart from
each other.  In the original version of the algorithm, the many-body
problem was solved by determining at each iteration of the simulation
a maximum time step such that each particle could interact with at
most one other particle during that time step
\cite{VanZon05,VanZon05_2}. In the enhanced version of the algorithm
presented here, called eGFRD, spherical protective domains are put
around single and pairs of particles \cite{Opplestrup:2006ta}. This
allows for an exact, asynchronous event-driven algorithm (see {\em
  Supporting Information}).

\subsection{MAPK model}
The model of the distributive, MAP kinase dual phosphorylation cycle
is sketched in Fig. \ref{fig:mapk_scheme} 
and described by
Eqs. \ref{eq:K1}-\ref{eq:K5}. The rate constants are $k_{1} =
0.027\, {\rm nM^{-1} \cdot s^{-1}}$, $k_{2} =
1.35\, {\rm s^{-1}}$, $k_{3}= 1.5\, {\rm s^{-1}}$, $k_{4} = 0.056\, {\rm
  nM^{-1} \cdot s^{-1}}$, $k_{5} = 1.73\, {\rm s^{-1}}$, $k_{6} = 15.0\,
{\rm s^{-1}}$, $k_{7} = \ln 2 / \tau_{\rm rel}$. The protein
diameter $\sigma = 5\, {\rm nm}$. $k_1$ and $k_4$ are the
intrinsic association rates, which are the association rates for two
species in contact; $k_2$ and $k_5$ are the intrinsic
dissociation rates \cite{VanZon06}. While in the particle-based model
the diffusion of the particles is simulated explicitly, in the
mean-field model based on the ODE chemical rate equations, diffusion
is described implicitly by renormalizing the association and
dissociation rates \cite{VanZon06}: $1/k_{\rm on} = 1/k_{\rm a} +
1/k_{\rm D}$ and $1/k_{\rm off} = 1/k_{\rm d} + K_{\rm eq}/k_{\rm D}$,
where $k_{\rm on}$ and $k_{\rm off}$ are the renormalized association
and dissociation rates, respectively, $k_{\rm a}=k_1,k_4$ and $k_{\rm
  d}=k_2,k_5$ are the respective intrinsic association and
dissociation rates, $k_{\rm D} = 4 \pi \sigma D$ is the
diffusion-limited association rate, and $K_{\rm eq} = k_{\rm a}/k_{\rm
  d} = k_{\rm on}/k_{\rm off}$ is the equilibrium constant. The
particles were put in a cubic volume of $1\, {\rm \mu m}^3$ with periodic
boundary conditions. The total enzyme concentration $\rm
[KK] + [P]$ is $100\, {\rm nM}$ corresponding to 60 copies of molecules
in the volume, and the total substrate concentration $\rm [K] + [Kp] +
[Kpp]$ is $200\, {\rm nM}$ or 120 copies of molecules in Figs 3, 4
and 5, and $500\, {\rm nM}$ or 300 copies of molecules in Fig 6.  The
processive model consists of the following six reactions, sharing the
same rate constants as the distributive model: $\KK + \K
\overset{k_1, k_2}{\longleftrightarrow} \KKtK
\stackrel{k_3}{\rightarrow} \KKtKp \stackrel{k_6}{\rightarrow}\KK +
\Kpp, \ \ \Ptase + \Kpp \overset{k_{1}, k_{2}}{\longleftrightarrow}
\PtKpp \stackrel{k_{3}}{\rightarrow}\PtKp \stackrel{k_6}
       {\rightarrow}\Ptase + \K.  $


\begin{acknowledgments}
  KT conducted part of the research as a Human Frontier Science
  Program Cross-Disciplinary Fellow at the Molecular Sciences
  Institute.  We thank Marco Morelli, Jeroen van Zon, Boris Kholodenko,
  Tom Shimizu, Frank Bruggeman
  and Steven Andrews for useful discussions, Moriyoshi Koizumi for
  help in implementation, and Institute for Advanced Biosciences of
  Keio University for computing facility.  The work is part of the
  research program of the ``Stiching voor Fundamenteel Onderzoek der
  Materie (FOM)'', which is financially supported by the ``Nederlandse
  organisatie voor Wetenschappelijk Onderzoek (NWO)''.
\end{acknowledgments}

\bibliographystyle{pnas-bolker}

\begin{thebibliography}{10}
\bibitem{Footnote1}
The optimum is due to the fact that when a substrate molecule is bound
to an enzyme molecule, the probability $k_3/(k_2+k_3)$ that it is
modified by the enzyme molecule instead of dissociating from it,
slightly increases with decreasing diffusion constant (the
dissociation rate $k_2$ decreases with decreasing $D$).
\end{thebibliography}

\end{document}


\title{Supporting Information \\ 
{\it \large Spatio-temporal correlations can
    drastically change the response of the MAPK pathway}}

\author{K. Takahashi, S. T\u{a}nase-Nicola and P. R. ten Wolde}

\maketitle

\tableofcontents

\newpage

\section{Enhanced Green's Function Reaction Dynamics}

\subsection{Overview}
We present the enhanced Green's Function Reaction Dynamics (eGFRD)
simulation algorithm. We provide the concepts required to understand
the outline of the algorithm, but details on the algorithm, such as the
actual mathematical expression for the employed Green's functions,
other numerical procedures, and performance analyses, will be
given in a forthcoming publication \cite{eGFRD}.

We solve the many-body reaction-diffusion problem by decomposing it
into a set of many one body (single) and two body (pair) problems, for
which analytical solutions (Green's functions) exist.  In the original
version of the GFRD algorithm, the many-body problem was solved by
determining at each step of the simulation a maximum time step $\Delta
t$ such that each particle could interact with at most one other
particle during that time step. In practice, the maximum time step
$\Delta t$ was determined as follows: 1) an interaction sphere of
radius $H \sqrt{6 D_i \Delta t}$ is drawn around each particle, where
$D_i$ is the diffusion constant of the $i$-th particle, and $H$ is a
user-set error control parameter (usually 3 or 4); 2) the maximum
time step $\Delta t$ is then set by the requirement that each interaction
sphere overlaps at most with one other interaction sphere.
Subsequently, for each single particle and pair of particles a tentative
reaction time is drawn, after which all particles are propagated
simultaneously up to the smallest tentative reaction time, or
to the maximum time step if that is smaller than the smallest
tentative reaction time \cite{VanZon05,VanZon05_2}.  Although already
up to five orders more efficient than conventional reaction Brownian
Dynamics \cite{morelli2008_2} and also very accurate by its own right,
the original GFRD algorithm has three major drawbacks; 1) due to the
synchronous nature, the decomposition into one and two-body problems
has to happen at every simulation step; 2) all components in the
system are propagated according to the smallest tentative reaction
time, making the performance sub-optimum; 3) the decomposition into
single particles and pairs of particles involves cut-off distances,
which makes the algorithm inexact.  The systematic error is controlled
by the $H$ parameter, which determines the probability that during a
time step $\Delta t$ a particle travels a distance further than the
maximum distance set by the requirement that each particle can
interact with at most one other particle. This means that there is a
trade-off between performance and error.

In the current work, we overcome the drawbacks of the original GFRD
scheme by putting protective domains around single particles and pairs
of particles \cite{Opplestrup:2006ta}. In this scheme, the next event
of a domain can either be a reaction, or one particle leaving the
domain. The tentative exit time for the latter event is computed by
imposing an absorbing boundary condition on the surface of the
protective domain. This makes the algorithm exact, and allows for an
asynchronous event-driven algorithm.

In the following sections, we explain how the reaction-diffusion
problem in a spherical protective domain is solved for the one-body
(Single problem) and two-body case (Pair problem), and then describe
how the simulations of the different domains are integrated.

\subsection{Single particle events}
\begin{figure}
  \center
 \includegraphics[scale=.5]{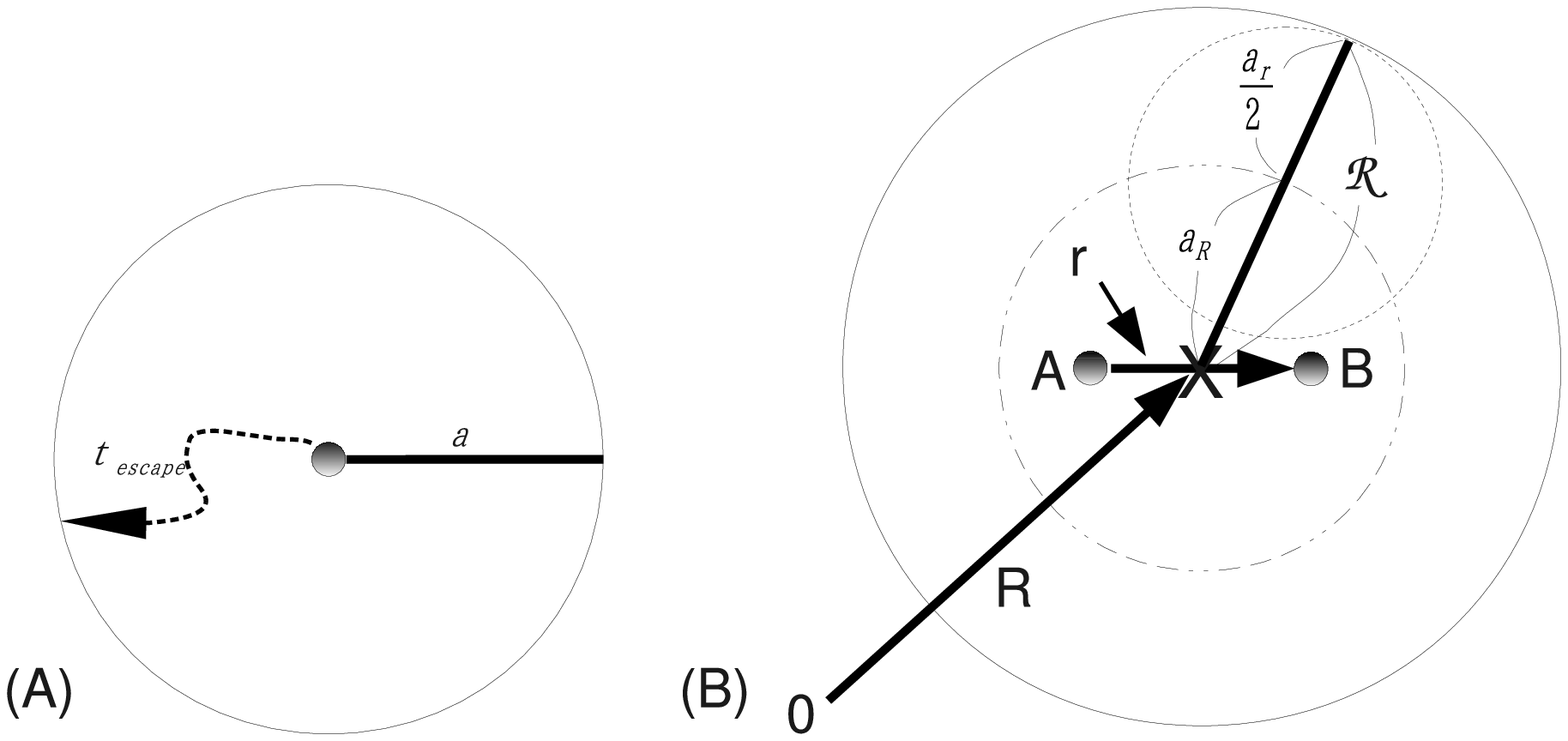}
 \caption{Single and Pair objects. To solve the many-body problem
   exactly, protective domains are put around single particles (A) and
   pairs of particles (B). (A) The radius of the protective domain of a
   single particle is denoted by $a$. (B) To solve the
   reaction-diffusion problem of two particles that can react with
   each other and diffuse within a protective domain with radius
   $\cal{R}$, we construct two protective domains: one for the
   center-of-mass ${\bf R}$, with radius $a_R$, and one for the
   inter-particle vector ${\bf r}$, with radius $a_r$. The radii $a_R$
   and $a_r$ can be freely chosen, provided that when ${\bf R}$ and
   ${\bf r}$ would reach their maximum lengths, i.e. when $|{\bf R}| =
   a_R$ and $|{\bf r}| = a_r$, the particles $A$ and $B$ would remain
   within the protective domain. The latter means that $a_R$ and $a_r$
   should satisfy the following two constraints: 1) $a_R +
   a_r D_A / (D_A + D_B) < {\cal R} -
   \sigma_A / 2$, which reflects that particle A should remain with
   the protective domain with radius $\cal R$; 2) $a_R +
   a_r D_B / (D_A + D_B) < {\cal R} -
   \sigma_B/2$, reflecting that $B$ should remain with the protective
   domain with radius $\cal R$. Although $a_R$ and $a_r$ can be freely
   chosen provided that these constraints are met, an efficient
   choice is given by $a_R^2 / D_R^2 = (a_r-{\bf r}_0)^2 / D_r^2$,
   meaning that the average time for ${\bf R}$ to reach the boundary
   of its domain by free diffusion, equals that of ${\bf r}$. To
   illustrate the constraints, panel B shows a scenario where ${\bf
     R}$ and ${\bf r}$ reach their maximum lengths; here,
   $D_A = D_B$.}
 \label{fig:egfrd}
\end{figure}

We consider a single particle of diameter $d$ surrounded by a
spherical protective shell of radius $a$ (Fig. \ref{fig:egfrd}(A)).
Motion of a freely diffusing spherical particle is described
by the Einstein diffusion equation,
\begin{eqnarray}
\partial_t p_1({\bf r}, t|{\bf r_0},t_0) =
D \nabla^2 p_1({\bf r}, t|{\bf r_0},t_0),  \label{eq:p1}
\end{eqnarray}
where the Green's function $p_1({\bf r}, t|{\bf r_0},t_0)$ denotes the
probability that the particle is at position ${\bf r}$ at time $t$
given that it was at ${\bf r_0}$ at time $t_0$.  We obtain the Green's
function $p_1({\bf r}, t|{\bf r_0},t_0)$ by solving the diffusion
equation, Eq. \ref{eq:p1}, with
the following initial and boundary conditions
\begin{eqnarray}
p_1({\bf r}, t_0|{\bf r_0},t_0) &=& \delta({\bf r} - {\bf r_0}), \label{eq:p1_ini}\\
p_1(|{\bf r-r_0}|=a, t|{\bf r_0},t_0) &=& 0, \label{eq:p1_bc}
\end{eqnarray}
where $\delta$ denotes the Dirac delta function.

From the Green's function one can obtain the survival probability
\begin{eqnarray}
S_1(t|{\bf r_0},t_0) &=& \int_{{\rm |{\bf  r-r_0}|<a}} d{\bf r} p_1({\bf r}, t|{\bf r_0},t_0), \label{eq:S1}
\end{eqnarray}
which is the probability at time $t$ that the particle remains within
the protective sphere of radius $a$.  This is related to the
probability per unit time that the particle escapes the domain for the
first time,
\begin{equation}
q_1^{\rm escape}(t|{\bf r}_0,t_0) = -\partial S(t|{\bf r}_0,t_0)  / \partial t.
\end{equation}
Sampling from this escape-propensity function yields a tentative escape
time $t_{\rm escape}$.

It is also possible that the particle undergoes a unimolecular
reaction. The probability that the next reaction happens in an
infinitesimal time interval $t$ and $t + dt$ is \cite{Gillespie}
\begin{eqnarray}
q_1^{\rm reaction} (t|t_0) dt = k \exp( - k ( t - t_0 ) ) dt,  \label{eq:q1}
\end{eqnarray}
where $k$ is the first-order reaction rate. This distribution can be
used to obtain the next tentative reaction time $t_{\rm reaction}$.  

The next event time of a {\em Single} is given by the smallest of the
two tentative event times, namely,
\begin{equation}
t_{\rm single} = \min(t_{\rm escape}, t_{\rm reaction}). \label{eq:singletime}
\end{equation}

\subsection{Particle pair events}

To describe the diffusion and the reaction of a pair of particles, we use the
distribution function $p_2({\bf r}_A,{\bf r}_B,t|{\bf r}_{A0},{\bf
  r}_{B0},t_0)$, which gives the probability that the particles $A$
and $B$ are at positions ${\bf r}_{A}$ and ${\bf r}_{B}$ at time $t$,
given that they were at ${\bf r}_{A0}$ and ${\bf r}_{B0}$ at time
$t_0$. This distribution function satisfies for ${\bf |r|}\geq\sigma$,
where $\sigma = (d_A + d_B)/2$ is the cross-section with $d_A$ and $d_B$
the diameters of particles $A$ and $B$, respectively, the following diffusion equation:
\begin{eqnarray}
\label{eq:Smol}
\partial_t p_2({\bf r}_A,{\bf
r}_B,t|{\bf r}_{A0},{\bf r}_{B0},t_0) = 
[D_A \nabla^2_A + D_B \nabla^2_B] \ p_2({\bf r}_A,{\bf r}_B,t|{\bf r}_{A0},{\bf r}_{B0},t_0).
\end{eqnarray}
We aim to solve this equation for two particles that can react with
each other and diffuse within a protective domain.  To our knowledge,
it is impossible to solve this equation and obtain the Green's function
directly.  We therefore apply the following tricks.

First, we make a coordinate transformation
\begin{eqnarray}
{\bf R} &\equiv& \frac{D_B {\bf r}_A + D_A {\bf r}_B}{D_A + D_B},\label{eq:RAB}\\
{\bf r} &\equiv& {\bf r}_B - {\bf r}_A \label{eq:rAB},
\end{eqnarray}
and define the operators
\begin{eqnarray}
\nabla_{\bf R} &\equiv& \partial/\partial {\bf R},\\
\nabla_{\bf r} &\equiv& \partial/\partial {\bf r}.
\end{eqnarray}
Eq.~\ref{eq:Smol} can then be rewritten as:
\begin{eqnarray}
\label{eq:SmolRr}
 \partial_t p_2({\bf R},{\bf r},t|{\bf R}_{0},{\bf r}_{0},t_0) 
	&=& [D_R\nabla_{\bf R}^2 +D_r \nabla_{\bf r}^2] \  p_2({\bf R},{\bf r},t|{\bf R}_{0},{\bf r}_{0},t_0), 
\end{eqnarray}
where $D_R \equiv D_AD_B/(D_A+D_B)$ and $D_r \equiv D_A+D_B$. 
 This equation
describes two independent random processes, one for the inter-particle
vector ${\bf r}$ and another for the center-of-mass  vector ${\bf R}$.
This means that the distribution function $p_2({\bf r}_A,{\bf
r}_B,t|{\bf r}_{A0},{\bf r}_{B0},t_0)$ can be factorized as
$p_{2}^{\bf R}({\bf
R},t|{\bf R}_0,t_0)p_{2}^{\bf r}({\bf r},t|{\bf r}_0,t_0)$ and that the above
equation can be reduced to one diffusion equation for the
coordinate ${\bf R}$ and another for the coordinate ${\bf r}$:
\begin{eqnarray}
\partial_t p_{2}^{\bf R}({\bf R},t|{\bf R}_0,t_0) &=& D_R \nabla_{\bf R}^2  p_{2}^{\bf R}({\bf R},t|{\bf R}_0,t_0), \label{eq:pR}\\ 
\partial_t p_{2}^{\bf r}({\bf r},t|{\bf r}_0,t_0) &=& D_r \nabla_{\bf r}^2  p_{2}^{\bf r}({\bf r},t|{\bf r},t_0) \label{eq:pr}.
\end{eqnarray}

The crux is now to define one protective domain for the interparticle
vector ${\bf r}$, with radius $a_r$, and another for the
center-of-mass vector ${\bf R}$, with radius $a_R$ (see
Fig. \ref{fig:egfrd}(B)). These domains have to be chosen such that
when the inter-particle vector ${\bf r}$ and the center-of-mass vector
${\bf R}$ would reach their maximum lengths, given by $|{\bf
  r}|=a_r$ and $|{\bf R}|=a_R$, respectively, the particles $A$ and $B$
would  still be within the protective domain for the two particles.

The diffusion equation for the center-of-mass vector now has to be
solved with the boundary conditions
\begin{eqnarray}
p_2^{\bf R}({\bf R},t_0| {\bf R}_0,t_0) &=&  \delta ({\bf R}-{\bf R}_0),\label{eq:R_ini} \\
p_2^{\bf R}(|{\bf R-R_0}|=a_{R},t |{\bf R}_0,t_0) &=&  0. \label{eq:R_bc}
\end{eqnarray}
This problem, of the center-of-mass diffusing in its protective domain,
is similar to that of the single particle diffusing in a protective
domain as discussed in the previous section. From the corresponding
propensity function $q_2^{\bf R}(t|{\bf r_0})$ we can draw a tentative
time $t_R$ at which the center-of-mass leaves its protective domain.

The solution for the diffusion equation for the inter-particle vector
is less trivial, since it should take into account not only that the
inter-particle vector can leave its protective domain, but also that
the two particles can react with each other. This reaction is included
as an extra boundary condition, yielding the following boundary
conditions for the inter-particle vector ${\bf r}$:
\begin{eqnarray}
p_2^{\bf r}({\bf r},t_0| {\bf r}_0,t_0)  &=&  \delta ({\bf r}-{\bf r}_0),  \label{eq:r_ini}\\
p_2^{\bf r}(|{\bf r}|=a_{r},t |{\bf r}_0,t_0)   &=&   0, \label{eq:r_bc}\\
-j(\sigma, t|{\bf r_0}, t_0) \equiv 4\pi \sigma^2 D_r
\frac{\partial}{\partial r}  p_2^{\bf r}({\bf r},t |{\bf r}_0,t_0)|_{|{\bf r}| = \sigma} &=& k_a p_2^{\bf r}(|{\bf r}| = \sigma,t |{\bf r}_0,t_0), \label{eq:radbc}
\end{eqnarray}
 where $\partial /\partial r$ denotes a
derivative with respect to the inter-particle separation $r$.
Eq.
 \ref{eq:radbc} is the boundary condition that describes the
 possibility that $A$ and $B$ can react with a rate $k_a$ once they
 are in contact. Here, $j(\sigma,t|{\bf r}_0,t_0)$ is the  radial
flux of probability $p_2^{\bf r}({\bf r},t|{\bf r}_0,t_0)$ through the
``contact'' surface of area $4\pi \sigma^2$. This boundary condition,
also known as a {\em radiation} boundary condition~\cite{Agmon90},
states that this radial flux of probability equals the intrinsic rate
constant $k_a$
 times the probability that the particles $A$ and $B$ are in
contact.  In the limit $k_a \rightarrow \infty$, the radiation
boundary condition reduces to an {\em absorbing} boundary condition
$p_2^{\bf r}({\bf |r|}=\sigma,t|{\bf r}_0,t_0)=0$, while in the limit
$k_a \rightarrow 0$ the radiation boundary condition reduces to a {\em
  reflecting} boundary condition.

From the Green's function for the inter-particle vector ${\bf r}$,
$p_2^{\bf r}({\bf r},t|{\bf r}_0,t_0)$, we can obtain two important
quantities.  The first is the time $t_{\rm bimo}$ at which the
inter-particle vector crosses the reaction surface given by $|{\bf
  r}|=\sigma$---meaning that the particles $A$ and $B$ react with
each other---and the other is the time $t_{r}$ at which it ``escapes''
through the boundary of the protective domain given by $|{\bf r}|=a_r$.
  The time
at which the next event happens, be it a reaction or an escape, can be
obtained through the survival probability, which is  given by
\begin{equation}
\label{eq:Sa}
S_2^{\bf r}(t|{\bf r}_0,t_0) = \int_{\sigma \leq |{\bf r}| < a_r}
d{\bf r} p_2^{\bf r}({\bf r},t|{\bf
r}_0,t_0).
\end{equation}
The propensity function
$q_2^{\bf r}(t|{\bf r}_0,t_0)$, which is the probability that the next event
happens between time $t$ and $t+dt$, is related to the survival
probability by
\begin{equation}
\label{eq:nra}
q_2^{\bf r}(t|{\bf r}_0,t_0) \equiv -\frac{\partial S_2^{\bf r}(t|{\bf r}_0,t_0)}{\partial t}.
\end{equation} 
To know which of the two event types, reaction or escape, happens at
time $t$, we split this quantity into two components,
\begin{eqnarray}
q_2^{\bf r} (t|{\bf r}_0,t_0) =& q^{\sigma}_2 (t|{\bf r}_0,t_0) + q^{a_r}_2
(t|{\bf r}_0,t_0) \\
 =& \int_{|{\bf r}|=\sigma} dS D_r
\frac{\partial}{\partial r} p_2^{\bf r}({\bf r},t|{\bf r}_0,t_0) -
\int_{|{\bf r}|=a_r} dS D_r
\frac{\partial}{\partial r} p_2^{\bf r}({\bf r},t|{\bf r}_0,t_0), \label{eq:q2r}
\end{eqnarray}
where in the first term $dS$ denotes an integral over the reaction
surface at $|{\bf r}|=\sigma$, and in the second term an integral over
the boundary of the protective domain $|{\bf r}|=a_r$.  The reaction
rate $q^{\sigma}_2(t|{\bf r}_0,t_0)$ is the probability that the {\em
  next} reaction for a pair of particles, initially separated by ${\bf
  r}_0$, occurs at time $t$, while the escape rate $q^{a_r}_2(t|{\bf
  r}_0,t_0)$ yields the probability that the inter-particle distance
reaches $a_r$ and escapes from the protective domain for the first
time at time $t$. We can draw the tentative time $t$ for the next event,
be it an escape or a reaction event, from Eq. \ref{eq:nra}, and then
determine which of the two takes place from the ratio of $q^{\sigma}_2
(t|{\bf r}_0,t_0)$ and $q^{a_r}_2(t|{\bf r}_0,t_0)$ at time
$t$. Alternatively, we can draw a tentative time for a bimolecular
reaction, $t_{\rm bimo}$, from $q^{\sigma}_2 (t|{\bf r}_0,t_0)$ and
a tentative time for an escape event, $t_r$, from $q^{a_r}_2(t|{\bf
  r}_0,t_0)$; which of the two events can occur is then the
one with the smallest tentative time (see below).  The function
$\frac{\partial}{\partial r}p^{\bf r}_2({\bf r},t|{\bf r}_0,t_0)$ can
be used to sample the exit points on the relevant surfaces.

It is possible that the particles $A$ and $B$ do not only react
with each other, but also can undergo a unimolecular reaction of the
type $X \rightarrow \dots$. 
In the same way as in the {\em Single} problem (Eq. \ref{eq:q1}), we can
also draw the times $t_{{\rm mono},A}$ and $t_{{\rm mono},B}$ at
which the particles $A$ and $B$ undergo a first-order reaction, respectively.

The next event of a pair of particles in a protective domain is thus
one of the following events: 1) the center-of-mass leaving its domain;
2) the inter-particle event leaving its domain; 3) a bimolecular
reaction; 4) unimolecular reaction of molecule A; 5) a unimolecular
reaction of molecule B. The event that actually takes place is the one
with the smallest tentative time. The next event time for a protective
domain with two particles is thus given by
\begin{equation}
  t_{\rm pair} = \min( t_{R}, t_{r}, t_{\rm bimo},t_{{\rm mono},A},
t_{{\rm mono},B}) \label{eq:pairtime}.
\end{equation}

\subsection{Algorithm outline}

The outline of the eGFRD algorithm is given by:

\begin{enumerate}
\item {\em Initialize:} Reset the simulator time ($t_{\rm sim}
  \leftarrow 0$).  For each particle in the system, draw a spherical
  protective domain of appropriate size.  When two particles are very
  close, create a {\em Pair} between them.  Otherwise, create a {\em
    Single} object for each of the particles.  Then, for each of the
  Single and the Pair objects, draw the next event type and the next event
  time according to the formulations in the previous sections, and
  chronologically order the events in the scheduler.

\item {\em Step:}  Pick the next event with the smallest scheduled time $t$
  from the scheduler.  Update the simulator time $t_{\rm sim} \leftarrow t$.

  \begin{itemize}
    \item {\em Single event}
      \begin{itemize}

        \item If the event is a {\em Single escape} event, then (1)
          propagate the particle to a randomly determined exit point
          on the surface of the protective domain; (2) check if there
          are protective domains that are close to the new position of
          the particle; (3) if there are, burst the neighboring
          domains, and propagate the particles in the burst domains
          to a new position, and check if the current {\em Single}
          particle can form a {\em Pair} with one of the neighboring
          particles;  (4) if a {\em Pair} is formed, discard the
          current {\em Single}, determine the new {\em Pair} event
          time (Eq. \ref{eq:pairtime}), and schedule the new {\em
            Pair} event on the scheduler; (5) for each of the
          particles contained in the stepping Single or the burst
          domains that are not used in formation of the {\em Pair}, draw a
          new domain and schedule a {\em Single} event on the
          scheduler.

        \item If the event type is {\em Single reaction}, (1)
          propagate the particle to a point ${\bf r}$ within the
          protective domain according to the {\em Single} Green's
          function $p_1({\bf r}, t_{\rm sim}|{\bf r_{\rm
              last}},t_{\rm last})$, where $t_{\rm last}$ is the
          time the Single was created or the last time it stepped, and
          ${\bf r_{\rm last}}$ is the position of the particle at
          $t_{\rm last}$; (2) execute the reaction by replacing
          the particle with one or more of the product particles
          placed next to each other; (3) for each of the newly
          created particles, draw a new protective domain and schedule
          a {\em Single event} on the scheduler.
      \end{itemize}

    \item {\em Pair event}
      \begin{itemize}
        \item If the event type is {\em Pair reaction}, meaning that
          the two particles in the domain react, (1) draw
          the new ${\bf R}$ from $p_2^{\bf R}({\bf R},t_{\rm sim}|
          {\bf R}_{\rm last},t_{\rm last})$, where ${\bf R}_{\rm
            last}$ is the position of ${\bf R}$ at $t_{\rm last}$, which
          is the time at which the {\em Pair} was formed;
          (2) remove particles A and B of the {\em Pair} from the
          simulator; (3) place the product particle(s) at the
          new position ${\bf R}$; (4) draw protective domain(s)
          around the new particle(s), and schedule {\em Single} event(s)
          on the scheduler.

        \item If the event type was {\em ${\bf r}$ escape}, meaning that the
          inter-article vector ${\bf r}$ leaves its protective domain,
          (1) sample the new ${\bf R}$ position as above; (2) sample
          the ${\bf r}$ exit point from $\frac{\partial}{\partial r}
          p^{\bf r}_2$; (3) determine the new positions of A and B,
          ${\bf r}_{\rm A}$ and ${\bf r}_{\rm B}$, by
          putting the ${\bf R}$ as calculated in (1), and the exit
          point ${\bf r}$ on the surface of the inter-particle
          protective domain as calculated in (2), into Eqs
          \ref{eq:RAB} and \ref{eq:rAB}; (4) delete the {\em Pair};
          (5) create a {\em Single} domain and schedule a new {\em
            Single} event on the scheduler for both A and B.

        \item If the event type is {\em ${\bf R}$ escape}, meaning that the
          center-of-mass leaves its domain, (1) sample the new
          inter-particle vector ${\bf r}$ with $p_{2}^{\bf r}({\bf
            r},t_{\rm sim}|{\bf r}_{\rm last},t_{\rm last})$, where
          ${\bf r}_{\rm last}$ is the inter-particle vector at the
          time $t_{\rm last}$ it was last updated; (2) sample the
          ${\bf R}$ exit point from $\frac{\partial}{\partial r}
          p^{\bf R}_2$; (3) displace the particles A and B to the new
          positions; (4) delete the {\em Pair}; (5) create two {\em
            Single}s and schedule them on the scheduler.

        \item If the event type was {\em single reaction}, (1) burst
          the pair domain and update the positions of the particles A and B
          by sampling $p_2^{\bf R}({\bf R},t_{\rm sim}| {\bf
            R}_{\rm last},t_{\rm last})$ and $p_{2}^{\bf
            r}({\bf r},t_{\rm sim}|{\bf r}_{\rm last},t_{\rm
            last})$; (2) execute the reaction of the reacting particle
          according to the same procedures as used in the {\em Single} event; (3)
          create a new {\em Single} domain for the other (non-reacting)
          particle and schedule it on the scheduler.

      \end{itemize}

  \end{itemize}
  
\item Go to 2.

\end{enumerate}

It can happen that more than two particles come very close to each
other, making it difficult to draw protective domains of sufficient
size around each of the particles\cite{Opplestrup:2006ta}; this could
bring the simulations to a standstill. To preempt this scenario, the
algorithm puts one protective domain around the ``squeezed'' particles
to form a third type of object called {\em Multi}. The particles in
this domain are propagated according to Brownian
dynamics\cite{morelli2008_2} until the particles recover from the
squeezed condition.  Since it is guaranteed that Brownian dynamics
converges to the correct solution when a sufficiently small step size is
used\cite{morelli2008_2}, this squeezing recovery procedure does not
affect the overall accuracy of the simulation.

The actual forms of the single and pair Green's functions, efficient
numerical evaluation methods for the Green's functions, more details
on the algorithm including the recovery procedure from squeezing, and
handling of surfaces will be described in a forthcoming
publication \cite{eGFRD}.

\newpage

\section{Rebinding-time  distribution}

In this section we present scaling relations for the
rebinding-time distributions of two particles that can diffuse and
react with each other in a large compartment.  These results give a
mathematical interpretation of the non-monotonic form of the
enzyme-substrate rebinding-time distributions, shown in Fig. 4 (panels
A and B) of the main text.

The problem is reduced to solving the reaction-diffusion equation for
a random walker that can diffuse in a domain internally bounded by a
sphere of radius $\sigma$, and to which it can bind with an intrinsic
rate $k_{\rm a}$ once it is in contact with the sphere. This
represents the evolution of the inter-particle vector ${\bf r}$
describing the distance between a substrate molecule and the enzyme
molecule that has just modified it.

The rebinding probability can be obtained from the Green's function
$p_2({\bf r},t|{\bf r}_0,t_0)$. The probability that a particle that
starts at the origin given by $|{\bf r}|=\sigma$ returns to the origin
at a later time $t$, is given by
\begin{equation}
 p_2(\sigma,t|\sigma,0)=\frac{\sigma-e^{D t/\sigma^2(1+h
     \sigma)^2}\sqrt{\pi D t}(1+h \sigma){\rm
     erfc}\left((1+h\sigma)\sqrt{\frac{D t}{\sigma^2}}\right)}{4 \pi
   \sigma^3 \sqrt{\pi D t}},
\label{eq:rebind}
\end{equation}
where ${\rm erfc}$ is the complementary error function. If we assume
that the enzyme becomes active immediately after enzyme-substrate
dissociation, then, according to Eq. \ref{eq:radbc}, the
rebinding-time probability distribution can be expressed as
\begin{equation}
p_{\rm reb}^0(t)=k_{\rm a} p_2(\sigma,t|\sigma,0).
\end{equation}

This rebinding-time distribution has a number of properties. Firstly,
the total probability that there is a rebinding is smaller than one:
\begin{equation}
  \int_0^\infty \! \! dt \, p_{\rm reb}(t)=\frac{1}{1+\frac{4 \pi D
      \sigma}{k_{\rm a}}} < 1.
\end{equation}
Secondly, upon a variable change $t=\tau
\frac{\sigma^2}{D(1+k_{\rm a}/\left(4 \pi D \sigma)\right)^2}$, the rebinding
probability distribution can be rescaled as
\begin{equation}
 p_{\rm reb}(\tau)=\frac{1}{1+\frac{4 \pi D \sigma}{k_{\rm a}}}f(\tau), \quad f(\tau)=\left[\frac{1}{\sqrt{\pi \tau}}-e^{\tau}{\rm erfc}\left(\sqrt{\tau}\right)\right] \label{eq:reb}.
\end{equation}
The function $f(\tau)$ has the shape of two power laws
$f(\tau)\simeq\frac{1}{\sqrt{\tau}}$ for $\tau\ll 1$ and
$f(\tau)\simeq\frac{1}{\tau \sqrt{\tau}}$ for $\tau\gg 1$, in
accordance with the results presented in Figure 2 of the main text. In
fact, we can estimate the time for the inflection point to be
$\tau_{\rm mol}=\frac{\sigma^2}{D(1+k_{\rm a}/\left(4 \pi D
    \sigma)\right)^2}$.  Here, $\tau_{\rm mol}$ represents the time
after which most of the rebindings correspond to particles that start
at contact, but wander away from the reaction sphere before they
return to and rebind the reaction sphere. We stress that these
trajectories are {\em rebinding} trajectories: we thus exclude
trajectories where particles diffuse in the bulk and come back in a
memory-less fashion (see also below). The $t^{-1/2}$ scaling for $t <
\tau_{\rm mol}$ can be understood by noticing that on this time scale
particles stay close to the surface of the reaction sphere; indeed, on
this time scale the particles essentially see a flat reaction surface,
meaning that the return-time distribution is that of a 1D random
walker as described in the main text. The $t^{-3/2}$ scaling for $t >
\tau_{\rm mol}$ can be understood by observing that on this time scale
the particles have diffused away from the surface of the sphere; the
particles now see the entire sphere, which means that the
rebinding-time distribution is that of a 3D random walker returning to
the origin. Interestingly, $\tau_{\rm mol}$ depends on $k_{\rm
  a}$. When $k_{\rm a}$ is increased, the probability that a particle
binds the target upon contact increases. Hence, the probability that
after a time $t$ the particle is still performing a 1D random walk
close to the surface, decreases as $k_{\rm a}$ increases---the
particle has either reacted with the surface, or escaped from the
surface, thus performing a 3D random walk.

  So far we have assumed that upon dissociation, the enzyme and
  substrate can rebind as soon as they are in contact
  again. If, however, they can only rebind after the enzyme has become
  active again, then the rebinding-time distribution is given by
\begin{equation}
 p_{\rm reb}(t)=\int_0^t \! d t' \, \int_{\sigma}^{\infty}\! \! 4 \pi
 r^2 dr \,p^*_2(r,t'|\sigma,0) k_{\rm a} p_{\rm act}(t') p_2(\sigma ,t-
 t'|r ,0),
\label{eq:p_reb}
\end{equation} 
where 
\begin{equation}
 p_{\rm act}(t)=k_{\rm act}e^{-k_{\rm act} t}
\end{equation}
is the enzyme reactivation distribution with $k_{\rm act} \equiv 1 /
\tau_{\rm rel}$,
and $p^*_2(r,t|r_0,0)$ is the solution of the Smoluchowski equation
with a reflecting boundary condition---this reflects the idea that when
the enzyme has not become active yet, the substrate cannot bind it.

While we do not know an analytical expression for the rebinding-time
distribution of Eq. \ref{eq:p_reb}, we can derive a lower bound
$p_{\rm reb}^{\rm min}(t)$ and an upper bound $p_{\rm reb}^{{\rm
    max}}(t)$ for it, such that $p_{\rm reb}^{\rm min}(t) \leq p_{\rm
  reb}(t) \leq p_{\rm reb}^{\rm max} (t)$. The upper bound $p_{\rm reb}^{\rm
  max} (t)$ is based on the inequality
\begin{equation}
p_2(\sigma ,t- t'|r ,0)\leq p_2(\sigma ,t- t'|\sigma ,0),
\end{equation}
with equality for $r=\sigma$. This inequality expresses the fact
that the probability that the particles are in contact
at a later time $t$ 
decreases with the initial distance. This yields the upper bound
\begin{equation}
 p_{\rm reb}^{\rm max}(t) = \int_0^t  \! d t' \,  k_{\rm a} p_{\rm act}(t') p_2(\sigma ,t- t'|\sigma ,0).
\label{eq:bound_up}
\end{equation}
Using the solution for $p_2(\sigma ,t- t'|\sigma ,0)$ as described in
\ref{eq:reb} one can show that for small $t$ the bound increases with
$t$ as $\sqrt{t}$, while for large $t$ it decreases with $t$ as
$\frac{1}{t \sqrt{t}}$.  The upper bound $p_{\rm reb}^{\rm max}(t)$
thus has a non-monotonic behavior, going to zero at both zero and
infinity. The position of the maximum depends on the two time scales
$\tau_{\rm rel}=1/k_{\rm act}$ and $\tau_{\rm mol}$ and is located in
the interval $\left[{\rm MIN}(\tau_{\rm mol}, \tau_{\rm rel}),{\rm
    MAX}(\tau_{\rm mol}, \tau_{\rm rel})\right]$.

The lower bound $p_{\rm reb}^{\rm min}(t)$ is based on the inequality
$p_2 (r,t|\sigma,0) \leq p_2^*(r,t|\sigma,0)$. This reflects the idea
that with a radiation boundary condition, the particle can react with
the reactive sphere (and thus leak out of the system), while with a reflecting boundary condition it
cannot. This yields the following expression for the lower bound
\begin{eqnarray}
 p_{\rm reb}^{\rm min}(t)&=&\int_0^t  \! d t' \,  k_{\rm a} p_{\rm
   act}(t') p_2(\sigma ,t|\sigma ,0),\\
 &=& (1-e^{-k_{\rm act}t}) k_a p_2(\sigma,t|\sigma,0),\\
&=& s_{\rm act} (t) p_{\rm reb}^0(t),
\label{eq:bound_low}
\end{eqnarray}
where $s_{\rm act}(t)$ is the probability that the enzyme is active
after time $t$ and $p_{\rm reb}^0(t)$ is the enzyme-substrate rebinding-time
distribution assuming that the enzyme is active at all times.
This lower bound has a number of interesting scaling regimes,
depending on the relative values of $\tau_{\rm rel}$ and $\tau_{\rm
  mol}$. They can be understood intuitively by making the
following observations: as discussed above, for
$t < \tau_{\rm mol}$, $p_{\rm reb}^0(t) \sim t^{-1/2}$, while for $t >
\tau_{\rm mol}$, $p_{\rm reb}^0(t) \sim t^{-3/2}$; moreover, for $ t
\ll \tau_{\rm rel}$, $s_{\rm act} (t) \sim t$, while for $t \gg
\tau_{\rm rel}$, $s_{\rm
  act}(t) \to 1$. Hence, for $t < {\rm MIN}(\tau_{\rm mol}, \tau_{\rm
  rel})$, $p_{\rm reb}^{\rm min}(t) \sim t^{-1/2} \times t \sim t^{+1/2}$, while
for $ t > {\rm MAX}(\tau_{\rm mol}, \tau_{\rm rel})$, $p_{\rm
  reb}^{\rm min}(t)
\sim t^{-3/2}$. Moreover, when $\tau_{\rm rel} < \tau_{\rm mol}$,
$p_{\rm reb}^{\rm min}(t) \sim t^{-1/2}$ for $\tau_{\rm rel} < t <
\tau_{\rm mol}$, because $s_{\rm act} \to 1$ and $p_{\rm reb}^0(t)
\sim t^{-1/2}$. And in the scenario that $\tau_{\rm mol} < \tau_{\rm
  rel}$, $p_{\rm reb}^{\rm min}(t) \sim t^{-1/2}$ for $\tau_{\rm mol}
< t < \tau_{\rm rel}$, because $s_{\rm act} \sim t$ and
$p_{\rm reb}^0(t)\sim t^{-3/2}$.

\begin{figure}[t]
\center
\includegraphics[width=12cm]{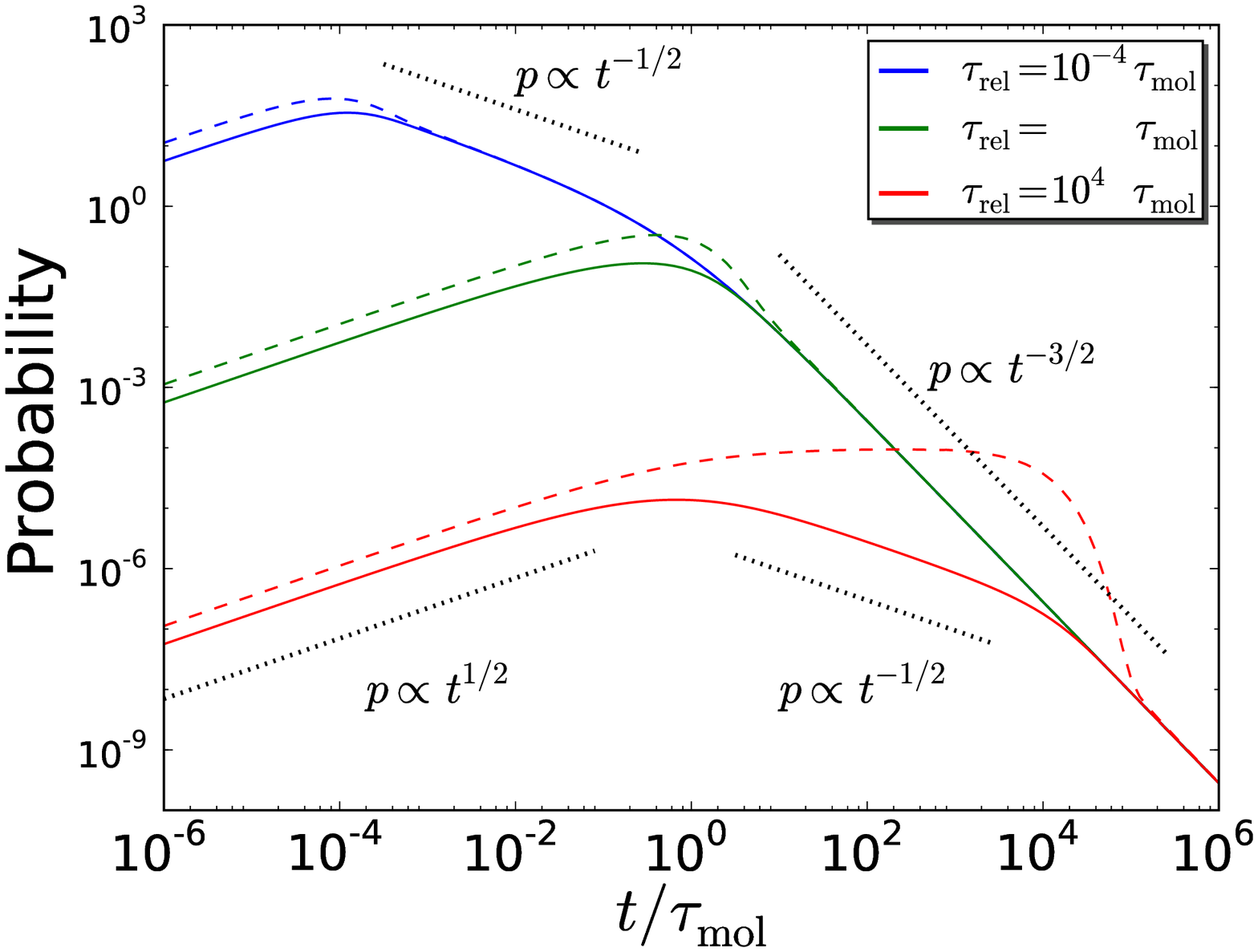}
\caption{The upper bound Eq. \ref{eq:bound_up} (dashed lines) and
  lower bound Eq. \ref{eq:bound_low} (solid lines) of the
  rebinding-time distribution, given by Eq. \ref{eq:p_reb}, for three
  different scenarios: 1) $\tau_{\rm rel} \ll \tau_{\rm mol}$ (blue
  lines); $\tau_{\rm rel} \approx \tau_{\rm mol}$ (green lines);
  $\tau_{\rm rel} \gg \tau_{\rm mol}$ (red lines). It is seen that
  both bounds converge when $t > \tau_{\rm rel}$. It is also seen that
  the difference between the bounds is rather small when $t <
  \tau_{\rm mol}$. The difference between the upper and lower bounds
  arises for $\tau_{\rm mol} < t < \tau_{\rm rel}$, when $\tau_{\rm
    rel} \gg \tau_{\rm mol}$ (red lines). This is because in this case
  the reactive sphere ({\em i.e.}, enzyme) is mostly still inactive,
  and the (substrate) particle thus tends to diffuse away from
  it. This phenomenon is captured by the lower bound, but not by the
  upper bound. For a comparison with the simulation results, see
  Fig. \ref{fig:prebbound_GFRD}.
  \label{fig:bound_up_low}}
\end{figure}

\begin{figure}[t]
{\bf (A)}\includegraphics[width=7.5cm]{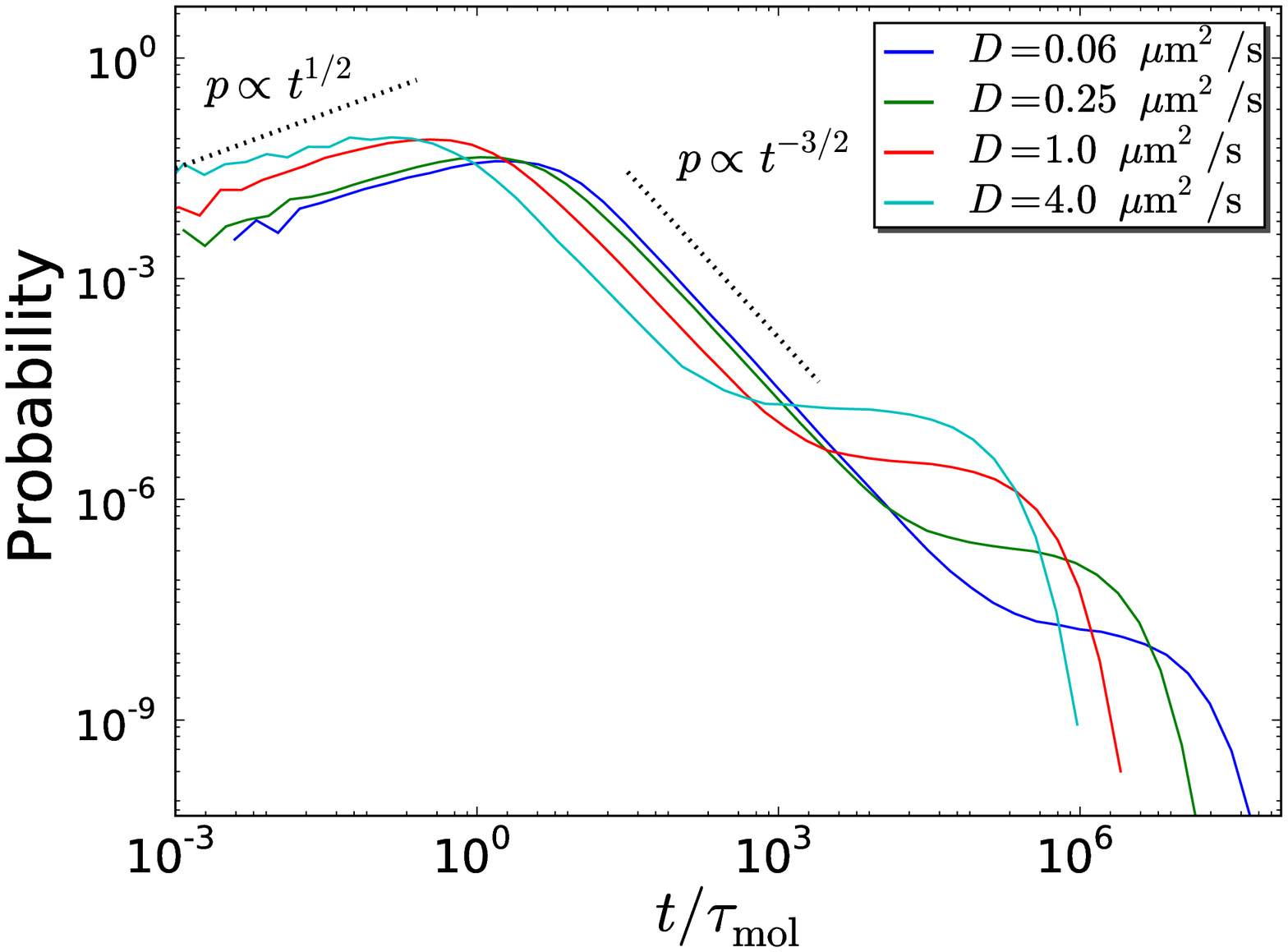}
{\bf (B)}\includegraphics[width=7.5cm]{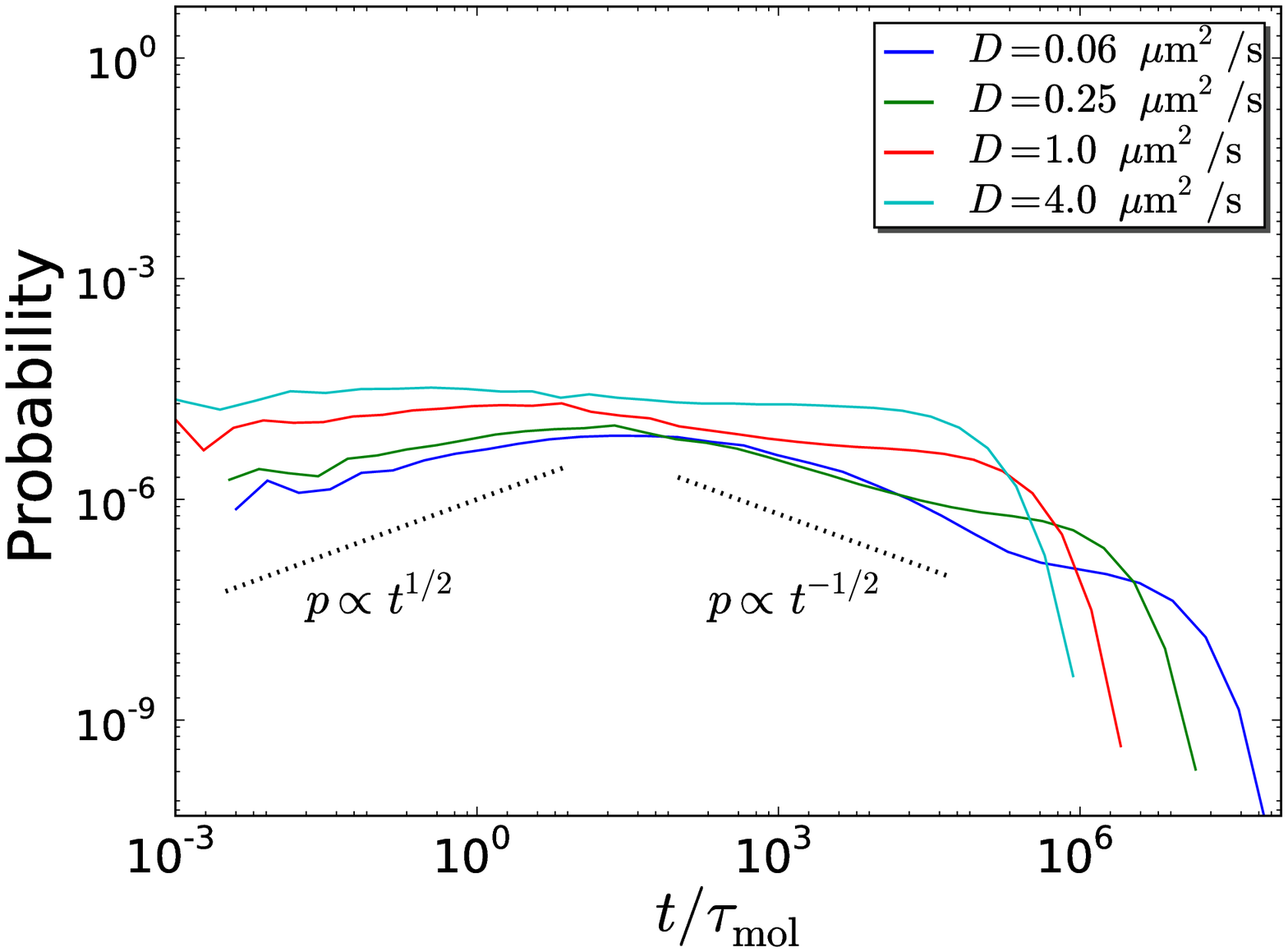}
\caption{The enzyme-substrate association-time distribution of Fig. 4
  of the main text, together with the scaling regimes as predicted by
  the analysis of the upper and lower bounds for the rebinding-time
  distribution (see Fig. \ref{fig:bound_up_low}); in panel A
  $\tau_{\rm rel} \approx \tau_{\rm mol}$, while in panel B $\tau_{\rm
    rel} \gg \tau_{\rm mol}$. For $t > \tau_{\rm
    bulk}$, the association-time distribution
  is exponential, because on this time scale the particles meet each
  other at random in the bulk. As predicted by the analysis  of the upper and lower
  bounds for the rebinding-time distribution (see
  Fig. \ref{fig:bound_up_low}),  the enzyme-substrate association-time
  distribution scales as $t^{1/2}$  for $t < {\rm MIN}(\tau_{\rm
    mol},\tau_{\rm rel})$, and as $t^{-3/2}$ for   ${\rm MAX}(\tau_{\rm
    mol},\tau_{\rm rel}) < t < \tau_{\rm bulk}$; while the $t^{1/2}$ scaling is
  seen in both panels, the $t^{-3/2}$ is only seen in panel A, because in
  panel B $\tau_{\rm bulk}$ approaches $\tau_{\rm rel}$. Panel B
   shows that 
the lower bound Eq. \ref{eq:bound_low}
   correctly predicts the $t^{-1/2}$ scaling for $\tau_{\rm mol} < t <
   \tau_{\rm rel}$, when $\tau_{\rm rel} \gg \tau_{\rm mol}$.
   \label{fig:prebbound_GFRD}}
\end{figure}

Fig. \ref{fig:bound_up_low} shows the predictions of the upper bound
Eq. \ref{eq:bound_up} and lower bound Eq. \ref{eq:bound_low} for the
rebinding-time distribution given by Eq. \ref{eq:p_reb}, for three
different scenarios: 1) $\tau_{\rm rel} \ll \tau_{\rm mol}$ (blue
line); 2) $\tau_{\rm rel} \approx \tau_{\rm mol}$ (green line); 3)
$\tau_{\rm rel} > \tau_{\rm mol}$ (red line). It is seen that in all 3
scenarios the upper and lower bound for $p_{\rm reb}(t)$ converge for
$t > \tau_{\rm rel}$. Indeed, in this regime, where the enzyme is
active, $p_{\rm reb}(t)$ scales as $t^{-3/2}$. It is also observed
that when $\tau_{\rm rel} \leq \tau_{\rm mol}$ (blue and green lines),
the difference between the upper and lower bound for $p_{\rm reb}(t)$
is very small, even when $t < \tau_{\rm rel}$. This implies that both
bounds are good approximations for $p_{\rm reb}(t)$; we can thus
conclude that, to a good approximation, $p_{\rm reb}(t)$ 
scales as $t^{1/2}$ for $t < {\rm MIN}(t_{\rm rel},\tau_{\rm mol})$ and
$t^{-1/2}$ for $\tau_{\rm rel} < t < \tau_{\rm mol}$ when $\tau_{\rm
  rel} < \tau_{\rm mol}$. The difference between the bounds arises
when $\tau_{\rm rel} > \tau_{\rm mol}$ (red line); in this scenario,
the bounds differ in the regime $\tau_{\rm mol} < t < \tau_{\rm
  rel}$. The question arises which bound is closer to the actual
rebinding-time distribution, $p_{\rm reb}(t)$. To this end, we compare
our analytical results with the simulation data, shown in
Fig. \ref{fig:prebbound_GFRD}.

In Fig. \ref{fig:prebbound_GFRD} we show the simulation results of
Fig.4 of the main text. When $t \geq \tau_{\rm bulk}$ the
rebinding-time distribution is exponential. This is due to particles
that come from the bulk, and bind the reactive sphere in a memory-less
fashion. This regime is not described by the analysis discussed above,
which is performed for the geometry of an infinite, spherical domain
internally bounded by a reactive sphere.  For this geometry, in three
dimensions, there is a probability that the particle escapes to
infinity without returning to and reacting with the reactive
sphere. In a bounded domain, when a particle escapes from the vicinity
of the reactive sphere into the bulk, it will return to the reactive
sphere on a time scale $\tau_{\rm bulk}$.  Therefore, the
rebinding-time distributions for the infinite domain analyzed above
and the finite domain of the simulations, are the same, but only up to
$\tau_{\rm bulk}$. This also means that to observe the different
power-law scaling behaviours, $\tau_{\rm bulk} \gg {\rm
  MAX}(\tau_{\rm rel},\tau_{\rm mol})$.

In panel A of Fig. \ref{fig:prebbound_GFRD}, $\tau_{\rm rel} \approx
\tau_{\rm mol}$, while in panel B $\tau_{\rm rel} \gg \tau_{\rm
  mol}$. In both scenarios $p_{\rm reb}(t) \sim (t)^{1/2}$ when $t <
{\rm MIN} (\tau_{\rm rel}, \tau_{\rm mol})$, in accordance with the
analysis of the upper and lower bounds of $p_{\rm reb}(t)$ presented
above. Both panels also show that when ${\rm MAX}(\tau_{\rm
  mol},\tau_{\rm rel}) < t < \tau_{\rm bulk}$, $p_{\rm reb}(t) \sim
t^{-3/2}$. An interesting regime is $\tau_{\rm mol} < t < \tau_{\rm
  rel}$ in the case that $\tau_{\rm rel} > \tau_{\rm mol}$ (panel
B). It is seen that the simulation results suggest that $p_{\rm
  reb}(t) \sim t^{-1/2}$ in this regime. This is predicted by the
lower bound for $p_{\rm reb}(t)$, Eq. \ref{eq:bound_low}, but not by
the upper bound, Eq. \ref{eq:bound_up} (see
Fig. \ref{fig:bound_up_low}). This can be understood by noting that in
this regime, $\tau_{\rm mol} < t < \tau_{\rm rel}$, the enzyme is
mostly still inactive and the particle can thus diffuse away from the
reactive sphere; while the lower bound of Eq. \ref{eq:bound_low}
captures this effect, the upper bound of Eq. \ref{eq:bound_up} does
not, since it is based on the inequality $p_2(\sigma ,t- t'|r ,0)\leq
p_2(\sigma ,t- t'|\sigma ,0)$.

\newpage

\section{The effect of concentration}
\begin{figure}[h]
\center
{\bf (A)}\includegraphics[width=6cm]{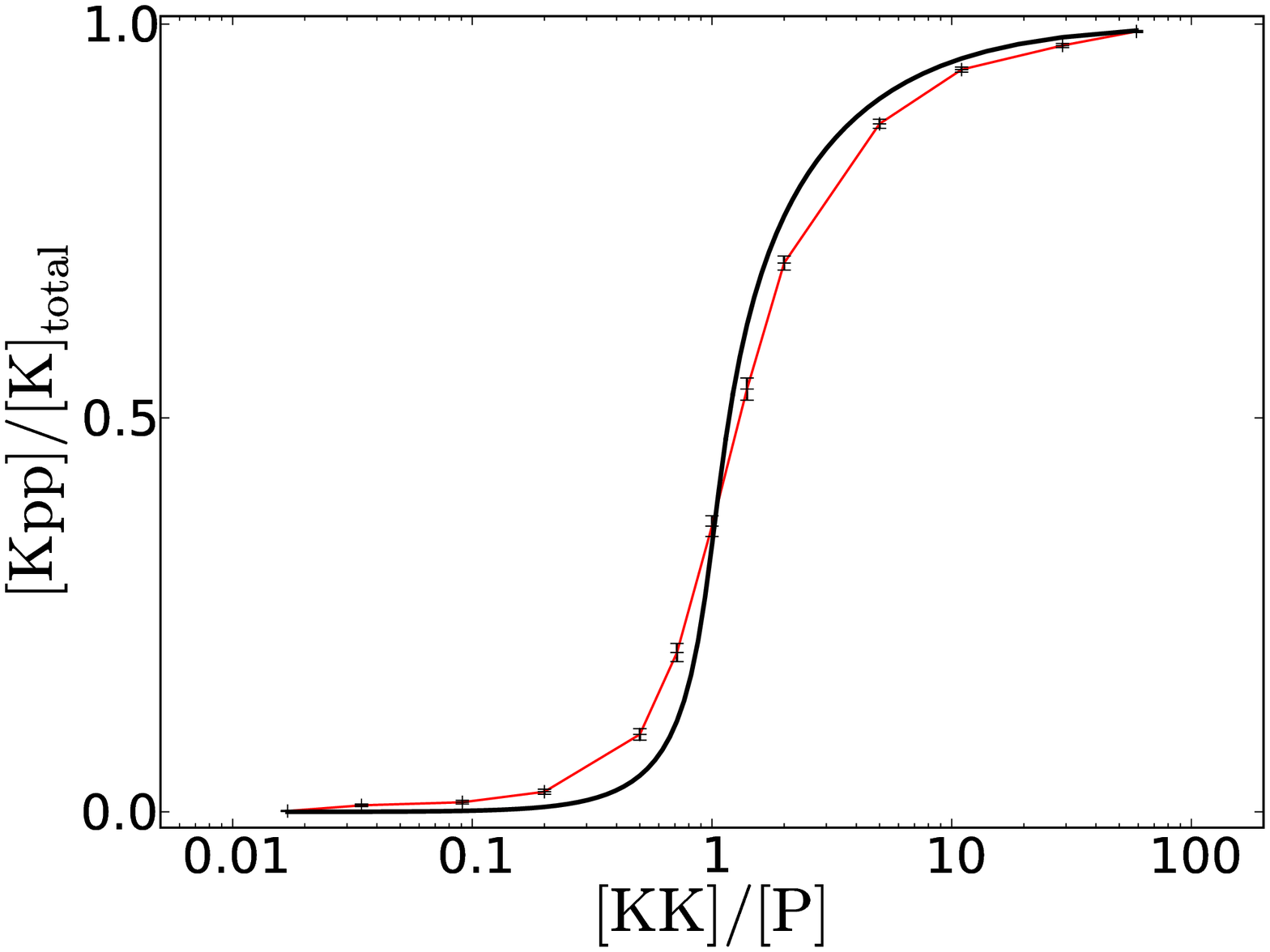}
{\bf (B)}\includegraphics[width=6cm]{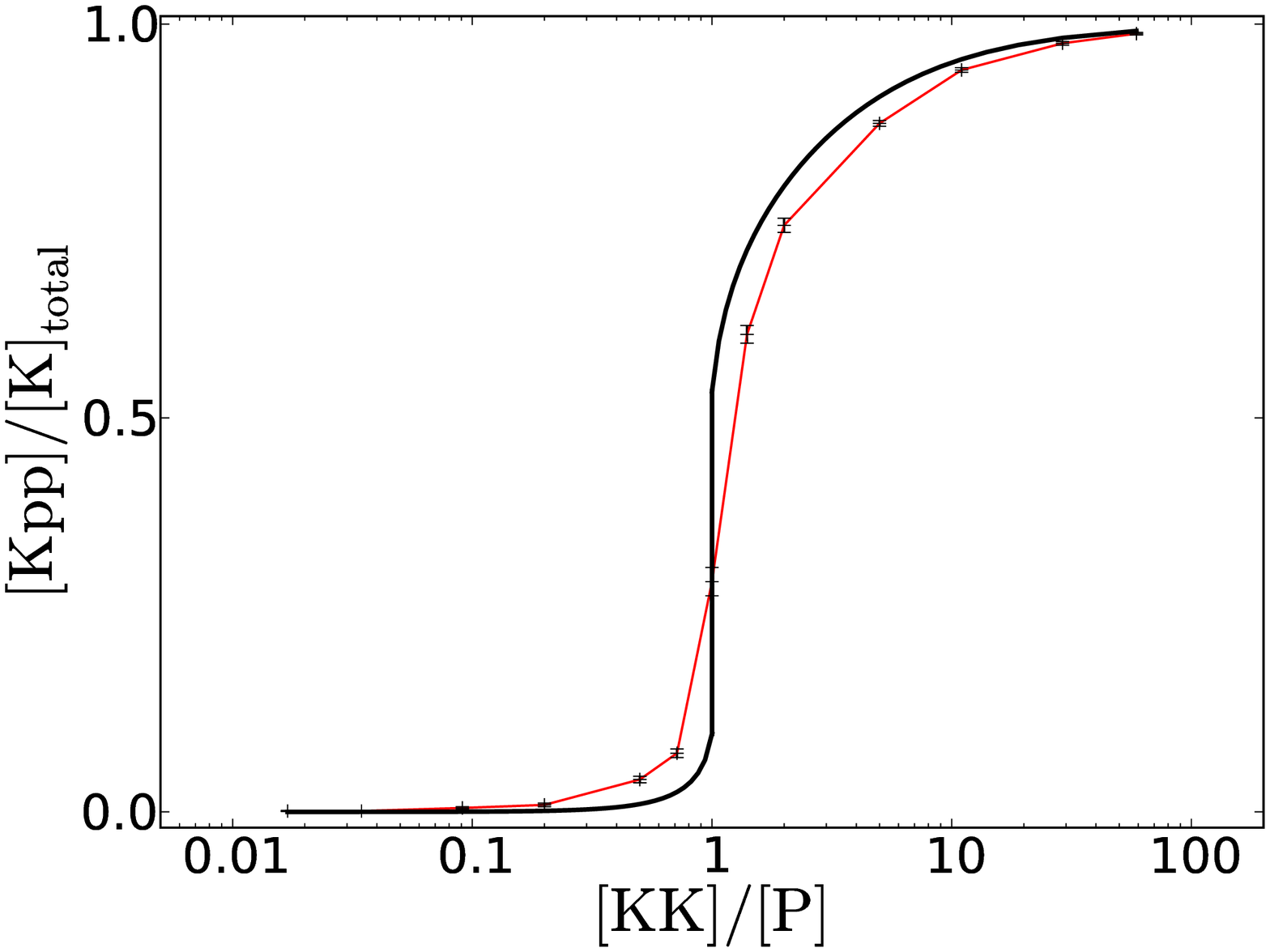}\\
{\bf (C)}\includegraphics[width=6cm]{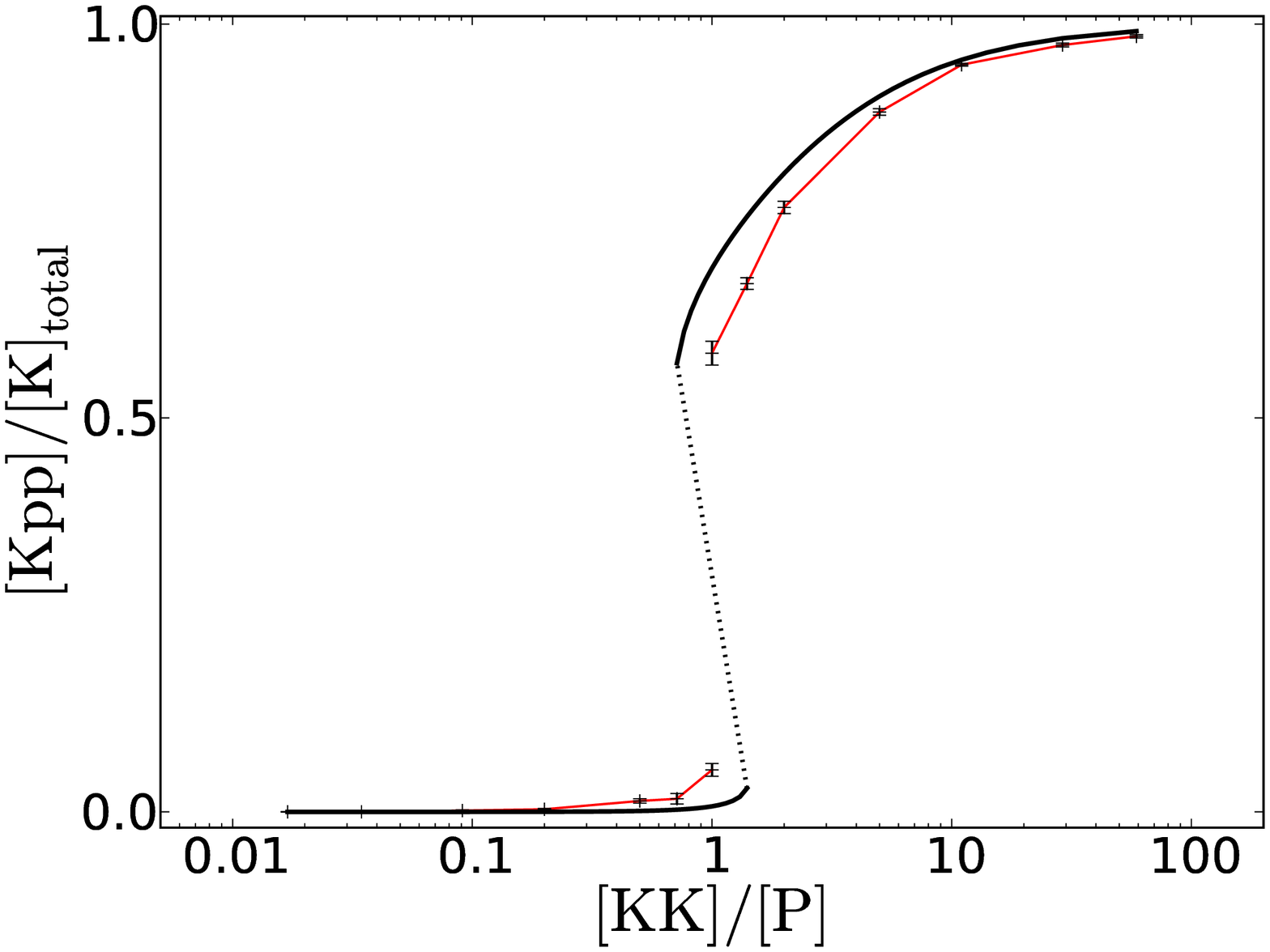}
{\bf (D)}\includegraphics[width=6cm]{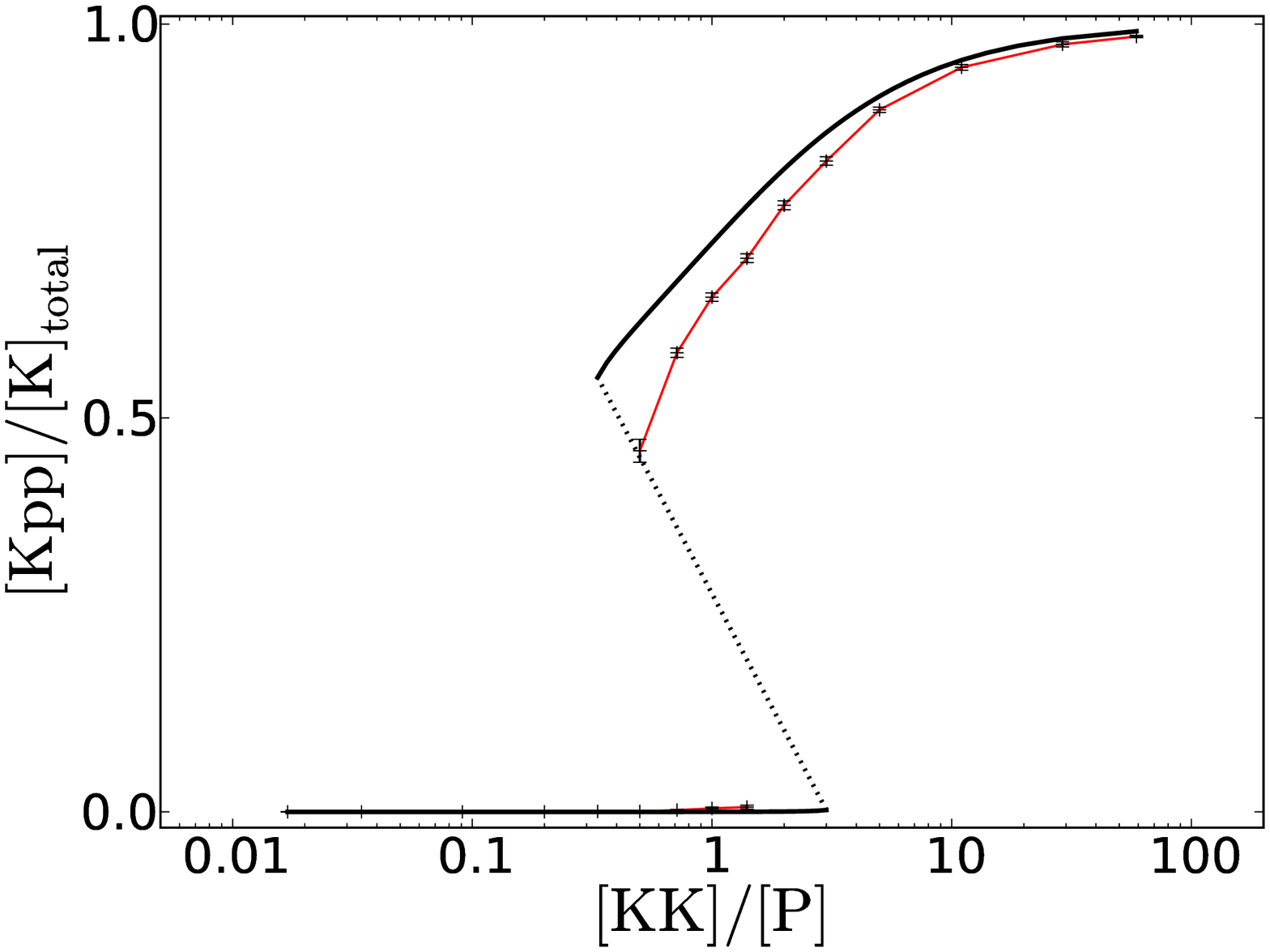}
\caption{Steady-state input-output relations for different
  concentrations.  The concentrations of all components are increased
  by the same factor. (A) Baseline parameter values; the
  concentrations equal those corresponding to Fig. 5(A) in the main
  text: $[\K]_{\rm total}= 200 {\rm nM}$, $[\KK]+[\Ptase]=100 {\rm
    nM}$).  (B) 3x concentration ($[\K]_{\rm total}= 600 {\rm nM}$,
  $[\KK]+[\Ptase]=300 {\rm nM}$), (C) 10x concentration ($[\K]_{\rm
    total}= 2 {\rm \mu M}$, $[\KK]+[\Ptase]=1 {\rm \mu M}$), (D) 100x
  concentration ($[\K]_{\rm total}= 20 {\rm \mu M}$,
  $[\KK]+[\Ptase]=10 {\rm \mu M}$).  
  \label{fig:ss_highconc}}
\end{figure}

Fig. \ref{fig:ss_highconc} shows the input-output relation as a
function of concentration. Here, the concentrations of all components
are increased by the same factor from the base-line values used in
Fig.5A of the main text. It is seen that both the particle-based model
and the mean-field model predict that an increase in concentration
induces bistability, although the concentration at which the
bifurcation occurs is higher in the particle-based model.

\begin{figure}[h]
\center
\includegraphics[width=7cm]{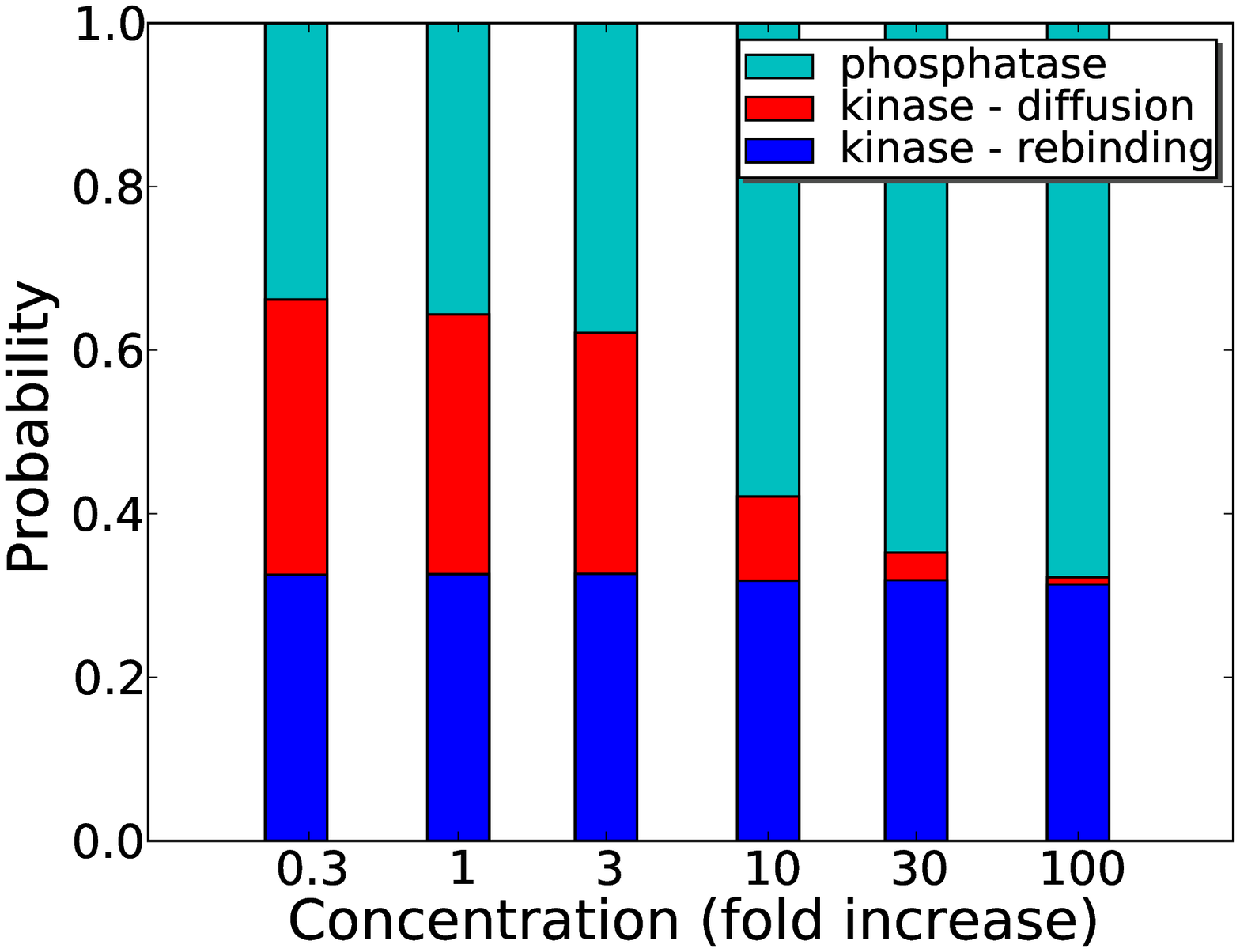}
\caption{The probability that a substrate molecule that has been
  phosphorylated once, will bind the same kinase molecule (blue),
  another kinase molecule (red) or a phosphatase molecule (green), for
  difference concentrations; the baseline values correspond to  Fig.5A of the main text and
  Fig. \ref{fig:ss_highconc}.  It is seen that the fraction of events
  where the substrate molecule binds the same kinase molecule again is
  fairly constant, while the fraction of events in which the substrate
  molecule binds another kinase molecule strongly drops in favor of
   those in which the substrate molecule binds a phosphatase
  molecule, when the concentration is increased by a factor 10 from
  the baseline value---as shown in 
  Fig. \ref{fig:ss_highconc}, the system now becomes bistable.
\label{fig:conc_events}}
\end{figure}

Fig. \ref{fig:conc_events} elucidates the origin of why an increase in
concentration can induce bistability, both in the mean-field model and
the particle-based model. Bistability arises when a substrate molecule
that has been phosphorylated once, is more likely to be
dephosphorylated again than to become fully phosphorylated (similarly,
the probability that after a fully phosphorylated molecule has been
dephosphorylated once becomes fully phosphorylated again, should be
higher than that it becomes fully dephosphorylated). We therefore plot
in Fig. \ref{fig:conc_events} the probability that a substrate that
has just been phosphorylated once, either binds the same kinase
molecule as the one that just phosphorylated it (this is most likely
due to a rebinding event), another kinase molecule (from the bulk), or
a phosphatase molecule (from the bulk); the system is in a state where
most substrate molecules are unphosphorylated.  It is seen that the
fraction of rebindings is fairly constant. This can be understood as
follows: 1) the probability that a molecule returns to the origin
before it looses memory where it came from is independent of the
concentration (see Fig.2 of the main text)---only the memory-less
returns from the bulk depend on concentration; 2) when a rebinding
event happens, it happens very fast: as Fig. 2 of the main text shows,
rebindings are dominated by events that occur on time scales of $ t <
1 \, {\rm ms}$. These time scales are so short, that the probability
that an enzyme molecule from the bulk interferes with a rebinding
event, is negligible, even up to concentrations of 100-1000 times the
baseline value, i.e. $10-100 \, \mu{\rm M}$; only above that concentration
can molecules from the bulk effectively compete with those undergoing
a rebinding trajectory, and will the probability that a dissociated
molecule rebinds drop significantly.  Up to a concentration of $10-100
\mu{\rm M}$, there is thus an essentially constant probability,
independent of the concentration, that both sites of a substrate
molecule are modified by the same enzyme molecule.  Now bistability
can arise when the antagonistic enzyme in the bulk wins the
competition from the agonistic enzyme undergoing the rebinding event
and the other agonistic enzymes in the
bulk. Fig. \ref{fig:conc_events} shows that when the concentration is
increased, the competition between the kinase (the agonist) in the
bulk and the phosphatase (antagonist) in the bulk changes in favor of
the phosphatase. This is because the system is in a state where the
substrate molecules are mostly unphosphorylated, and in this state the
kinase molecules become increasingly sequestered by the
unphosphorylated substrate molecules as the concentration is
increased. This increases the probability that a molecule that has
just been phosphorylated once, will bind a phosphatase (antagonist),
which will drive it back towards the unphosphorylated
state. Increasing the overall concentration thus changes the
competition between the kinases and the phosphatases in the bulk in
such a way that the driving force towards a state in which the
substrate molecules are either fully unphosphorylated or fully
phosphorylated, increases. In essence, increasing the concentration
drives the system deeper into the bistable regime. This makes it
possible to overcome the effect of rebindings, which tends to drive
the system out of the bistable regime, as shown in Fig. 6 of the main
text.

\bibliographystyle{pnas-bolker}